\documentclass[10pt,journal,epsfig]{IEEEtran}
\usepackage[dvips]{graphicx}
\usepackage{graphicx}
\usepackage{amssymb}
\usepackage{cite}
\usepackage{subfigure}
\usepackage{stfloats}
\usepackage{amsmath}
\newcounter{MYtempeqncnt}

\usepackage{subeqnarray}
\usepackage{algorithm}
\usepackage{algpseudocode}
\usepackage{color}
\usepackage{cases,comment}
\usepackage[Symbol]{upgreek}
\makeatletter

\newcommand{\Rmnum}[1]{\expandafter\@slowromancap\romannumeral #1@}
\makeatother

\newtheorem{property}{Property}
\newtheorem{proposition}{Proposition}
\newtheorem{remark}{Remark}
\newtheorem{theorem}{Theorem}
\newtheorem{lemma}{Lemma}

\begin{document}
\IEEEoverridecommandlockouts
\title{On Artificial-Noise Aided Transmit Design for Multi-User MISO Systems with Integrated Services}
\author{Weidong Mei, Zhi Chen, \IEEEmembership{Senior Member, IEEE}, Lingxiang Li, Jun Fang \IEEEmembership{Member, IEEE} \\and Shaoqian Li \IEEEmembership{Fellow, IEEE}
\thanks{This work was supported in part by the National Natural Science Foundation of China under Grants 61631004 and 61571089.}
\thanks{The authors are with National Key Laboratory of Science and Technology on Communications, University of Electronic Science and Technology of China, Chengdu (611731), China (e-mails: mwduestc@gmail.com; chenzhi@uestc.edu.cn; LiLX@std.uestc.edu.cn; JunFang@uestc.edu.cn; lsq@uestc.edu.cn).
}}
\maketitle

\begin{abstract}
This paper considers artificial noise (AN)-aided transmit designs for multi-user MISO systems in the eyes of service integration. Specifically, we combine two sorts of services, and serve them simultaneously: one multicast message intended for all receivers and one confidential message intended for only one receiver. The confidential message is kept perfectly secure from all the unauthorized receivers. Our goal is to jointly design the optimal input covariances for the multicast message, confidential message and AN, such that the achievable secrecy rate region is maximized subject to the sum power constraint. This secrecy rate region maximization (SRRM) problem is a nonconvex vector maximization problem. To handle it, we reformulate the SRRM problem into a provably equivalent scalar optimization problem and propose a searching method to find all of its Pareto optimal points. The equivalent scalar optimization problem is identified as a secrecy rate maximization (SRM) problem with the quality of multicast service (QoMS) constraints. Further, we show that this equivalent QoMS-constrained SRM problem, albeit nonconvex, can be efficiently handled based on a two-stage optimization approach, including solving a sequence of semidefinite programs (SDPs). Moreover, we also extend the SRRM problem to an imperfect channel state information (CSI) case where a worst-case robust formulation is considered. In particular, while transmit beamforming is generally a suboptimal technique to the SRRM problem, we prove that it is optimal for the confidential message transmission whether in the perfect CSI scenario or in the imperfect CSI scenario. For implementation efficiency, we also analyze the computational complexity of our proposed methods and put forward two suboptimal schemes and two possible extensions. Finally, numerical results demonstrate that the AN-aided transmit designs are effective in expanding the achievable secrecy rate regions, and that the suboptimal strategies can achieve near-optimal performance.
\end{abstract}

\section{Introduction}
\IEEEPARstart{H}{igh} transmission rate and secure communication are basic demands for the future fifth-generation (5G) cellular networks. A heuristic way is to merge coexisting services, typically, multicast service and confidential service, into one integral service for one-time transmission, referred to as \emph{physical-layer service integration} (PHY-SI). Service integration is in fact not a new concept: traditional service integration techniques rely on upper-layer protocols to allocate different services on different logical channels, which is quite inefficient. On the contrary, PHY-SI enables coexisting services to share the same resources by exploiting the physical characteristics of wireless channels, thereby significantly increasing the spectral efficiency. The technique of PHY-SI could also find a wide range of applications in the commercial and military areas. For example, many commercial applications, e.g., advertisement, digital television, Internet telephony, and so on, are supposed to provide personalized service customization. As a consequence, confidential service and public service are collectively provided to satisfy the demand of different user groups. A crucial problem lies in how to establish the security of the confidential service while not compromising the public service. In battlefield scenarios, it is essential to propagate commands with different security levels to the frontline. The public information should be distributed to all soldiers, while the confidential information can only be accessed by specific soldiers.

The respective investigation on physical-layer multicasting and physical-layer security has received lots of attention in much literature. Herein we give a very brief review on relevant literature. Physical-layer multicasting offers a way to efficiently transmit common messages that all receivers can decode, and it is required that the rate successfully decoded by all users be maximized. Therefore, physical-layer multicasting strategies for instantaneous rate maximization have become the centerpiece of research activities, epitomized in \cite{jindal2006capacity,sidiropoulos2006transmit, kim2011optimal,zhu2012precoder,lee2013a,wu2013physical,du2013optimum}. Comparatively, due to the broadcast nature of wireless medium, physical layer security approach is playing an increasingly important role in wireless communication recently. It can achieve significant security performance without using secret keys whose distribution and management may lead to security vulnerability in wireless systems. Different transmit strategies against eavesdroppers have been developed with various levels of eavesdropper channel state information (ECSI) available to the transmitter; see existing surveys and tutorials \cite{shiu2011physical,he2013wireless,hong2013enhancing,mukherjee2014principles,schaefer2015secure,yener2015wireless, wang2015enhancing,liu2016physical} and the references therein. In the literature, artificial noise (AN)-aided transmission has been demonstrated as an effective way to combat eavesdroppers\cite{negi2005secret,liao2011qos,li2013transmit,zheng2015multi,zheng2017safe}. Recently, there is growing interest in an emerging topic in the area of physical-layer security, termed as \emph{confidential broadcasting} \cite{liu2009secrecy,liu2010multiple}. In this topic, a transmitter broadcasts multiple confidential messages to all receivers. Each confidential message is intended for one specified receiver but required to be perfectly secret from the others. Different approaches have been proposed in e.g., \cite{fakoorian2013on,park2016weighted,park2016secrecy} to maximize the sum secrecy rate under this system model.

Currently many research activities concentrated on PHY-SI from the viewpoint of information theory. In particular, Csisz{\'a}r and K{\"o}rner's work in \cite{csiszar1978broadcast} established the fundamental limit on the maximum rate region of PHY-SI that can be applied reliably under the secrecy constraint (i.e., the secrecy capacity region), where the optimal integration of multicast service and confidential service was derived in a discrete memoryless broadcast channel (DMBC). In \cite{Hung2010Multiple,liu2013new,ekrem2012capacity}, the authors extended the results to the case with multiple-input multiple-output (MIMO) Gaussian channels. Wyrembelski and Boche's work \cite{wyrembelski2014strong} deduced the achievable secrecy rate region under channel uncertainty in a compound broadcast channel, which makes it possible to seek the robust transmit strategies of PHY-SI. Furthermore, Wyrembelski and Boche amalgamated broadcast service, multicast service and confidential service in bidirectional relay networks \cite{Wyrembelski2012Physical}, in which a relay adds an additional multicast message for all nodes and a confidential message for only one node besides establishing the conventional bidirectional communication. However, the aforementioned works only aimed to derive capacity results or determine the existence of coding strategies that result in certain rate regions \cite{Schaefer2014Physical}. Such rate regions are always characterized by a union with regard to (w.r.t) all possible transmit covariance matrices subject to certain power constraints. For ease of practical implementation, especially in the multi-antenna wireless systems, it is also necessary to treat PHY-SI from the view point of signal processing, i.e., find the optimal input covariance matrices of the transmitted messages for maximizing the achievable secrecy rate regions. Such optimization problems turn out to be generally nonconvex, which also leads to the unsatisfying fact that most works on PHY-SI end when a certain characterization of a rate region is obtained.

In this paper, we handle the PHY-SI from the view point of signal processing, i.e., find the optimal input covariance matrices for the transmitted messages, with either perfect or imperfect CSI. Specifically, we consider the multiuser multiple-input single-output (MISO) broadcast channel (BC) with multiple receivers and two sorts of messages: a multicast message intended for all receivers, and a confidential message intended for merely one receiver. The confidential message must be kept perfectly secure from all other unauthorized receivers. To further enhance the security performance, we enable the transmitter to send artificial noise to degrade the reception at all unauthorized receivers. It follows that our considered system model is actually a generalization of that in physical-layer security. For example, in PHY-SI, the unauthorized receivers play a dual role. On the one hand, they are able to eavesdrop the confidential information deliberately, just as that in traditional physical-layer security. On the other hand, they are legitimate users in terms of the multicast service, and hence their quality of multicast service (QoMS) should be guaranteed above a certain threshold. As a result, the use of AN will fall into a dilemma: Excessive use of AN will degrade the QoMS at all receivers, while limited use of it cannot attain the best security performance. To the best of our knowledge, the only prior work tackling the transmitter optimization in the PHY-SI context is \cite{Hung2010Multiple}, where a reparameterizing method is proposed. However, this method is only applicable to a simple two-receiver MISO setting with perfect CSI. Moreover, this method itself involves solving a sequence of convex feasibility problems, which is computationally expensive to implement.

This paper aims to jointly optimize the input covariance matrices of the multicast message, confidential message and AN, to maximize the achievable secrecy rate region in a more general and convenient way. Our problem formulation considers multiple single-antenna unauthorized receivers, with perfect or imperfect CSI on the links of \emph{all} receivers. This secrecy rate region maximization (SRRM) problem turns out to be a biobjective vector optimization problem. Our goal is to find all Pareto optimal solutions of this SRRM problem. Unfortunately, the method of scalarization, a standard technique to seek Pareto optimal points of a vector optimization problem, might not yield all Pareto optimal solutions due to the non-convexity of our optimization problem \cite{boyd2009convex}. To deal with it, we degrade this vector optimization problem into an equivalent scalar one. Then it is proved that all Pareto optimal solutions of the primal SRRM problem can be efficiently exhausted by this means. Our main contributions are summarized as follows.
\begin{enumerate}
  \item For the perfect CSI case, we derive an equivalent scalar optimization problem to the primal SRRM problem by following the above-mentioned idea. Nonetheless, the equivalent problem still remains non-convex. To handle it, we first reformulate it as a two-stage optimization problem. Then it is shown that the outer problem can be handled by performing a one-dimensional search, while the inner problem is an SDP problem. Further, we extend the SRRM problem to an imperfect CSI case, where a worst-case robust formulation is considered. By adopting a similar way as that in the perfect CSI case, this worst-case SRRM problem could also be solved.
  \item For implementation efficiency, we first analyze the feasibility of transmit beamforming to achieve the obtained Pareto optimal performances, since the single-stream transmit beamforming requires lower implementation complexity than the high-rank transceiver schemes. It is proved that transmit beamforming is an optimal strategy for the confidential information transmission, which applies to the perfect CSI case as well as to the imperfect CSI case. In addition, we give complexity analysis of our proposed two-stage approach, and show that the resultant computational complexity is polynomial with regard to (w.r.t.) the problem size for achieving at least $\epsilon$-suboptimality, with either perfect or imperfect CSI. Furthermore, we propose two suboptimal schemes to implement PHY-SI with lower complexity and two possible extensions to show the scalability of our proposal.
  \item Finally, we examine the AN's efficacy from the numerical results. The numerical results demonstrate that in PHY-SI, AN could also enhance the overall security performance, as that in traditional physical layer security, without compromising the QoMS.
\end{enumerate}

This paper is organized as follows. Section \Rmnum{2} provides the system model description and problem formulation. The optimization aspects of our formulated designs are addressed in Section \Rmnum{3}, for the scenario with perfect CSI. Sections \Rmnum{4} describes extensions of our present work to the scenario with imperfect CSI. Section \Rmnum{5} introduces our proposed suboptimal PHY-SI schemes and possible extensions. The performance of the proposed transmit designs is studied using several simulation examples in Section \Rmnum{6}, and conclusions are drawn in Section \Rmnum{7}.

The notation of this paper is as follows. Bold symbols in capital letter and small letter denote matrices and vectors, respectively. ${{(\cdot)}^{H}}$, $\rm{rank}(\cdot)$ and $\text{Tr}(\cdot )$ represent conjugate transpose, rank and trace of a matrix, respectively. ${\mathbb{R}}_{+}$ and ${\mathbb{H}}_{+}^n$ denote the set of nonnegative real numbers and of $n$-by-$n$ Hermitian positive semidefinite (PSD) matrices. The $n \times n$ identity matrix is denoted by ${\mathbf{I}}_n$. $\mathbf{x}\sim \mathcal{CN}(\mathbf{\mu },\mathbf{\Omega })$ denotes that \textbf{x} is a complex circular Gaussian random vector with mean $\mathbf{\mu}$ and covariance $\mathbf{\Omega}$. $\mathbf{A}\succeq \mathbf{0}$ $(\mathbf{A}\succ \mathbf{0})$ implies that $\mathbf{A}$ is a Hermitian positive semidefinite (definite) matrix. ${\left\| \cdot \right\|}$ represents the vector Euclidean norm. $K$ represents a proper cone, and $K^*$ represents a dual cone associated with $K$.

\section{System Model and Problem Formulation}
We consider the downlink of a multiuser system in which a multi-antenna transmitter serves $K$ receivers, and each receiver has a single antenna. Assume that all receivers have ordered the multicast service and receiver 1 further ordered the confidential service\footnote{In this paper, we assume that only one receiver orders the confidential service within a single time slot. In practice, this corresponds to the case where the confidential service is provided to all receivers in a \emph{round-robin} manner to strengthen the security of confidential messages and to reduce the operational complexity at the transmitter.}. To enhance the security of the confidential service, the transmitter utilizes a fraction of its transmit power to send artificially generated noise to interfere the unauthorized receivers (eavesdroppers), i.e., receiver 2 to receiver $K$. To facilitate the description, let us denote ${\cal K} \buildrel \Delta \over = \{1,2,...,K\}$ and ${{\cal K}_e} \buildrel \Delta \over = {\cal K}/\{ 1\} $ as the indices of all receivers and of all unauthorized receivers, respectively.

\begin{figure}[!t]
\centering
\includegraphics[width=3in]{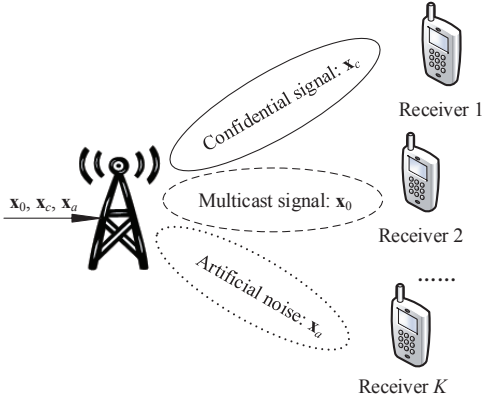}
\caption{Multiuser system model with integrated services}\label{sysmodel}
\end{figure}

The received signal at receiver $k$ is modeled as
\begin{equation}\label{yk}
  {y_k} = \;{{\bf{h}}_k}{\bf{x\;}} + \;{z_k}, k=1,2,\cdots,K
\end{equation}
respectively, where ${{\mathbf{h}}_k}\in {{\mathbb{C}}^{{1\times {N}_{t}}}}$ is the channel vector between the transmitter and receiver $k$, ${N}_{t}$ is the number of transmit antennas employed by the transmitter, and $z_k$ is independent identically distributed (i.i.d.) complex Gaussian noise with zero mean and unit variance. ${{\mathbf{x}}}\in {{\mathbb{C}}^{{{N}_{t}}}}$ is the transmitted signal vector which consists of three independent components, i.e.,
\begin{equation}\label{x3c}
  {\bf{x\;}}\; = \;{{\bf{x}}_0} + \;{{\bf{x}}_c} + \;{{\bf{x}}_a},
\end{equation}
where ${\bf{x}}_{0}$ is the multicast message intended for all receivers, ${\bf{x}}_{c}$ is the confidential message intended for receiver 1, and ${\bf{x}}_{a}$ is the artificial noise. We assume $\mathbf{x}_{0} \sim \mathcal{CN}(\mathbf{0},\mathbf{Q}_0)$, $\mathbf{x}_{c} \sim \mathcal{CN}(\mathbf{0},\mathbf{Q}_c)$ \cite{Hung2010Multiple}, where $\mathbf{Q}_0$ and $\mathbf{Q}_c$ are the transmit covariance matrices. The AN ${\bf{x}}_{a}$ follows a distribution $\mathbf{x}_{a} \sim \mathcal{CN}(\mathbf{0},\mathbf{Q}_a)$, where $\mathbf{Q}_a$ is the AN covariance. An exemplification of our system model is given in Fig.\,\ref{sysmodel}.

Denote $R_0$ and $R_c$ as the achievable rates associated with the multicast and confidential messages, respectively. Then an achievable secrecy rate region is given as the set of nonnegative rate pairs $(R_0,R_c)$ satisfying\footnote{We should point out that under the case where the secrecy rate is always zero, it is trivial to investigate the secrecy rate region, since the region would be degraded into a line segment on the axis of multicast rate. Thus, in this paper, we only focus on the nontrivial cases.} (cf. \cite{Hung2010Multiple,liang2009information})
\begin{subequations}\label{Region1}
\begin{align}
&{R_0} \le \mathop {\min }\limits_{k \in {\cal K}} {C_{m,k}}({{\bf{Q}}_0},{{\bf{Q}}_c},{{\bf{Q}}_a}),\\
&{R_c} \le {C_b}({{\bf{Q}}_c},{{\bf{Q}}_a})-\mathop {\max }\limits_{k \in {{\cal K}_e}} {C_{e,k}}({{\bf{Q}}_c},{{\bf{Q}}_a}),
\end{align}
\end{subequations}
where
\begin{subequations}\label{Region2}
\begin{align}
&{C_{m,k}}({{\bf{Q}}_0},{{\bf{Q}}_c},{{\bf{Q}}_a}) \buildrel \Delta \over = \log \left( {1 + \frac{{{{\bf{h}}_k}{{\bf{Q}}_0}{\bf{h}}_k^H}}{{1 + {{\bf{h}}_k}({{\bf{Q}}_c} + {{\bf{Q}}_a}){\bf{h}}_k^H}}} \right),\label{R2.1}\\
&{C_b}({{\bf{Q}}_c},{{\bf{Q}}_a}) \buildrel \Delta \over = \log \left( {1 + \frac{{{{\bf{h}}_1}{{\bf{Q}}_c}{\bf{h}}_1^H}}{{1 + {{\bf{h}}_1}{{\bf{Q}}_a}{\bf{h}}_1^H}}} \right),\\
&{C_{e,k}}({{\bf{Q}}_c},{{\bf{Q}}_a}) \buildrel \Delta \over = \log \left( {1 + \frac{{{{\bf{h}}_k}{{\bf{Q}}_c}{\bf{h}}_k^H}}{{1 + {{\bf{h}}_k}{{\bf{Q}}_a}{\bf{h}}_k^H}}} \right),
\end{align}
\end{subequations}
and $\text{Tr}(\mathbf{Q}_0+\mathbf{Q}_c+\mathbf{Q}_a) \le P$ with $P$ being total transmission power budget at the transmitter. ${C_{m,k}}$ is the achievable rate associated with the multicast message at receiver $k$, $C_b$ and $C_{e,k}$ are the mutual information at receiver 1 and the unauthorized receivers, respectively.

The secrecy rate region (\ref{Region1}) implies that all receivers first decode their common multicast message by treating the confidential message as noise, and then receiver 1 acquires a clean link for the transmission of its exclusive confidential message, where there is no interference from the multicast message. This can be achieved by following the same encoding schemes adopted in \cite{Hung2010Multiple}.

With perfect CSI being available at the transmitter, our work focuses on the design of $\mathbf{Q}_0$, $\mathbf{Q}_c$ and $\mathbf{Q}_a$, under an achievable SRRM formulation with power constraint. This problem is a vector maximization problem, with cone $K=K^*={\mathbb {R}}_ + ^2$, i.e.,
\begin{subequations}\label{op1}
\begin{align}
\nonumber&\mathop {\max}\limits_{{{\bf{Q}}_0},{{\bf{Q}}_a},{{\bf{Q}}_c},R_0,R_c} \left({\text{w.r.t.}}\; {\mathbb {R}}_ + ^2 \right)\; \left( {{R_0},{R_c}} \right)\\
\text{s.t.}\quad &\mathop {\min }\limits_{k \in {\cal K}} {C_{m,k}}({{\bf{Q}}_0},{{\bf{Q}}_c},{{\bf{Q}}_a}) \ge {R_0} ,\label{op1a}\\
& {C_b}({{\bf{Q}}_c},{{\bf{Q}}_a})-\mathop {\max }\limits_{k \in {{\cal K}_e}} {C_{e,k}}({{\bf{Q}}_c},{{\bf{Q}}_a}) \ge {R_c},\\
&\text{Tr}({{\bf{Q}}_0} + {{\bf{Q}}_a} + {{\bf{Q}}_c}) \le P,\\
&{{\bf{Q}}_0} \succeq {\bf{0}}, {{\bf{Q}}_a} \succeq {\bf{0}}, {{\bf{Q}}_c} \succeq {\bf{0}}.
\end{align}
\end{subequations}

\begin{remark}
Hereby we remark that it is valid to assume that the CSI on the links of all receivers and the number of unauthorized receivers are perfectly known at the transmitter in the PHY-SI. The reason is that all receivers have to register in the network for ordering the multicast service. During the registration or lease renewal, the receivers are required to feed their CSI back to the transmitter noiselessly, which could be achieved by utilizing a low-rate transmission with suitable quantization schemes \cite{love2008overview}. Nonetheless, considering the effect of channel aging, we will also investigate the case of imperfect channel knowledge at the transmitter in Section \Rmnum{4}.
\end{remark}

Substituting (\ref{Region2}) into (\ref{op1}), one can check that (\ref{op1}) is equivalent to the following vector optimization problem.
\begin{subequations}\label{op2}
\begin{align}
\nonumber&\mathop {\max}\limits_{{{\bf{Q}}_0},{{\bf{Q}}_a},{{\bf{Q}}_c},R_0,R_c} \left({\text{w.r.t.}}\; {\mathbb {R}}_ + ^2 \right)\; \left( {{R_0},{R_c}} \right)\\
\text{s.t.}\quad &\mathop {\min}\limits_{k \in {\cal K}}\;{\log \frac{{1 + {{\bf{h}}_k}({{\bf{Q}}_c} + {{\bf{Q}}_a} + {{\bf{Q}}_0}){\bf{h}}_k^H}}{{1 + {{\bf{h}}_k}({{\bf{Q}}_c} + {{\bf{Q}}_a}){\bf{h}}_k^H}}} \ge {R_0}, \label{op2a}\\
& \log \frac{{1 + {{(1 + {{\bf{h}}_1}{{\bf{Q}}_a}{\bf{h}}_1^H)}^{ - 1}}{{\bf{h}}_1}{{\bf{Q}}_c}{\bf{h}}_1^H}}{{\mathop {\max}\limits_{k \in {{\cal K}_e}} 1 + {{(1 + {{\bf{h}}_k}{{\bf{Q}}_a}{\bf{h}}_k^H)}^{ - 1}}{{\bf{h}}_k}{{\bf{Q}}_c}{\bf{h}}_k^H}} \ge {R_c},\\
&\text{Tr}({{\bf{Q}}_0} + {{\bf{Q}}_a} + {{\bf{Q}}_c}) \le P,\\
&{{\bf{Q}}_0} \succeq {\bf{0}}, {{\bf{Q}}_a} \succeq {\bf{0}}, {{\bf{Q}}_c} \succeq {\bf{0}}.
\end{align}
\end{subequations}
The SRRM problem (\ref{op2}) is a nonconvex vector optimization problem and thus difficult to solve. In the next section, we will elaborate our approaches to attacking (\ref{op2}).

\section{A Tractable Approach to the SRRM Problem}
A standard technique for dealing with the vector optimization problem is referred to as \emph{scalarization} \cite{boyd2009convex}. Its basic idea is to maximize the weighted sum of the two objectives, i.e., ${R_0}$ and ${R_c}$. By varying the weight vector, it could yield different maximal objective values, associated with \emph{Pareto optimal solutions} of the primal vector optimization problem. However, for a nonconvex vector optimization problem like (\ref{op1}), this method might not find all Pareto optimal points \cite{boyd2009convex}.

\subsection{An Equivalent Scalar Optimization Problem of (\ref{op2})}
In view of the limitation of the scalarization, now we develop another approach to find all Pareto optimal points of (\ref{op2}). Specifically, we first fix the variable $R_0$ as a constant $\tau _{ms} \ge 0$. As a result, the maximization of the vector $(R_0,R_c)$ will be degraded into the maximization of a scalar $R_c$, with the optimization problem given in (\ref{op3}). As it will be proved in Theorem \ref{theorem1}, by varying the parameter $\tau _{ms}$ and solving the problem (\ref{op3}), all Pareto optimal solutions of (\ref{op2}) can be found.
\begin{subequations}\label{op3}
\begin{align}
\nonumber {g^*}({\tau _{ms}})&= \mathop {\max}\limits_{{{\bf{Q}}_0},{{\bf{Q}}_a},{{\bf{Q}}_c}} \log \frac{{1 + {{(1 + {{\bf{h}}_1}{{\bf{Q}}_a}{\bf{h}}_1^H)}^{ - 1}}{{\bf{h}}_1}{{\bf{Q}}_c}{\bf{h}}_1^H}}{{\mathop {\max}\limits_{k \in {{\cal K}_e}} 1 + {{(1 + {{\bf{h}}_k}{{\bf{Q}}_a}{\bf{h}}_k^H)}^{ - 1}}{{\bf{h}}_k}{{\bf{Q}}_c}{\bf{h}}_k^H}}\\
\text{s.t.}\;\;&\mathop {\min}\limits_{k \in {\cal K}}{\log \frac{{1 + {{\bf{h}}_k}({{\bf{Q}}_c} + {{\bf{Q}}_a} + {{\bf{Q}}_0}){\bf{h}}_k^H}}{{1 + {{\bf{h}}_k}({{\bf{Q}}_c} + {{\bf{Q}}_a}){\bf{h}}_k^H}}} \ge \tau _{ms}, \label{op3a}\\
&\text{Tr}({{\bf{Q}}_0} + {{\bf{Q}}_a} + {{\bf{Q}}_c}) \le P,\\
&{{\bf{Q}}_0} \succeq {\bf{0}}, {{\bf{Q}}_a} \succeq {\bf{0}}, {{\bf{Q}}_c} \succeq {\bf{0}}.
\end{align}
\end{subequations}
In (\ref{op3}), the variable $R_c$ is discarded as a slack variable. It follows that $\tau _{ms}$ can be interpreted as preset requirement of the achievable multicast rate, and that (\ref{op3}) is an SRM problem with QoMS constraints. Actually, when we set $\tau _{ms}=0$, (\ref{op3}) becomes a conventional AN-aided SRM problem for multi-user MISO system. On the contrary, the confidential message transmission will be terminated provided that $\tau _{ms}$ is set above a threshold $\tau _{\max}$ given by
\begin{equation}\label{max_tau}
{\tau _{\max }} = \mathop {\max }\limits_{{{\bf{Q}}_0} \succeq {\bf{0}}, \text{Tr}({{\bf{Q}}_0}) \le P} \mathop {\min }\limits_{k \in \cal {K}} \log (1 + {{\bf{h}}_k}{{\bf{Q}}_0}{\bf{h}}_k^H).
\end{equation}
It is easy to find that $\tau _{\max}$ is the multicast capacity, and the optimization problem (\ref{max_tau}) can be solved via an SDP reformulation; see, e.g., \cite{jindal2006capacity, wu2013physical}.

Problem (\ref{op3}) is closely related to (\ref{op2}), and the crucial problem lies in whether problem (\ref{op3}) guarantees a complete inclusion of Pareto optimal solutions of problem (\ref{op2}).
\begin{theorem}\label{theorem1}
The rate pair $({\tau _{ms}},{g^*}({\tau _{ms}}))$ is a Pareto optimal point of (\ref{op2}), and all Pareto optimal points of (\ref{op2}) can be obtained by varying $\tau _{ms}$'s lying within $[0,\tau _{\max}]$.
\end{theorem}
\begin{IEEEproof}
First, we claim that problem (\ref{op3}) has some interesting properties as below, which will play a key role in the proof of Theorem \ref{theorem1}.
\begin{property}\label{property1}
The maximum objective value of problem (\ref{op3}) is obtained only when the equality in (\ref{op3a}) holds.
\end{property}
\begin{property}\label{property2}
The optimal objective value of (\ref{op3}), denoted as ${g^*}({\tau _{ms}})$, is monotonically decreasing w.r.t. $\tau _{ms}$.
\end{property}

The proof of Property \ref{property1} can be simply accomplished by contradiction: Assume the maximum value of problem (\ref{op3}) is obtained when the equality in (\ref{op3a}) does not hold, with ${{\mathbf{Q}}_a}$ unchanged, we multiply ${{\bf{Q}}_c}$ and ${{\bf{Q}}_0}$ by a scaling factor $\eta\;(\eta > 1)$ and $\xi\;(0 < \xi < 1)$, respectively, to equalize (\ref{op3a}) while keeping the total power constant. Then, we can always find a larger objective value for (\ref{op3}) in this way, which is contrary to the assumption.

Next we focus on the proof of Property \ref{property2}. Note that when $\tau_{ms}$ increases, the feasible region of problem (\ref{op3}) would be shrank. Thus, ${g^*}({\tau_{ms}})$ must be monotonically nonincreasing w.r.t $\tau_{ms}$. Furthermore, we claim that any two distinct $\tau_{ms}$ cannot generate an identical objective value of (\ref{op3}), since it will contradict Property \ref{property1}. This completes our proof of Property \ref{property2}.

Let us denote the set of objective values (1-by-2 vectors) of feasible points of (\ref{op2}) as $\cal O$. Then, we assume that there exist two different nonnegative rate pairs $(r_1,r_2), (r_3,r_4) \in \cal O$ for which ${r_1} \ne {r_3}$. From our problem formation of (\ref{op3}) and Property \ref{property1}, it is immediate to get $(r_1,{g^*}(r_1)) \succeq _{{\mathbb {R}}_ + ^2} (r_1,r_2), (r_3,{g^*}(r_3)) \succeq _{{\mathbb {R}}_ + ^2} (r_3,r_4)$. According to Property 2, if $r_1 \mathbin{\lower.3ex\hbox{$\buildrel>\over {\smash{\scriptstyle<}\vphantom{_x}}$}}  r_3$, then we will have ${g^*}(r_1) \mathbin{\lower.3ex\hbox{$\buildrel<\over {\smash{\scriptstyle>}\vphantom{_x}}$}} {g^*}(r_3)$. Consequently $(r_1,{g^*}(r_1))$ and $(r_3,{g^*}(r_3))$ are both Pareto optimal points of (\ref{op2}), since it is impossible to increase any one element of $(r_1,{g^*}(r_1))$ (resp. $(r_3,{g^*}(r_3)$) without decreasing the other one element of it. Substituting $r_1$ (or $r_3$) by $\tau _{ms}$, we then complete the proof.
\end{IEEEproof}

\begin{remark}
It should be mentioned that from the proof of Theorem \ref{theorem1}, $({\tau_{ms}},{g^*}({\tau_{ms}}))$ is also a boundary point of (\ref{Region1}). This implies that, in the specific context here, the Pareto optimal points of (\ref{op1}) are equivalent to the boundary points of (\ref{Region1}). When there is no ambiguity, the terms ``boundary points'' and ``Pareto optimal points'' will be used interchangeably in the following sections of this paper.
\end{remark}

\subsection{A Charnes-Cooper Transformation-Based Line Search Method for (\ref{op3})}
However, the equivalent QoMS-constrained SRM problem (\ref{op3}) still remains nonconvex. We now focus on deriving an SDP-based optimization approach for problem (\ref{op3}). To start with, we first rewrite (\ref{op3}) as
\begin{subequations}\label{op4}
\begin{align}
\nonumber&{g^*}(\tau ') = \mathop {\max}\limits_{{{\bf{Q}}_0},{{\bf{Q}}_a},{{\bf{Q}}_c},\alpha \ge 1} \log\left( {\frac{{1 + {{\bf{h}}_1}({{\bf{Q}}_c} + {{\bf{Q}}_a}){\bf{h}}_1^H}}{{\alpha (1 + {{\bf{h}}_1}{{\bf{Q}}_a}{\bf{h}}_1^H)}}} \right)\\
\text{s.t.}\;& \log \left(1 + \frac{{{{\bf{h}}_k}{{\bf{Q}}_c}{\bf{h}}_k^H}}{{1 + {{\bf{h}}_k}{{\bf{Q}}_a}{\bf{h}}_k^H}}\right) \le \log \alpha ,\forall k \in {{\cal K}_e},\label{op4a}\\
&{{\bf{h}}_k}{{\bf{Q}}_0}{\bf{h}}_k^H - \tau '{{\bf{h}}_k}({{\bf{Q}}_a} + {{\bf{Q}}_c}){\bf{h}}_k^H - \tau ' \ge 0,\forall k \in {\cal K}, \label{op4b}\\
&\text{Tr}({{\bf{Q}}_0} + {{\bf{Q}}_a} + {{\bf{Q}}_c}) \le P,\label{op4c}\\
&{{\bf{Q}}_0} \succeq {\bf{0}}, {{\bf{Q}}_a} \succeq {\bf{0}}, {{\bf{Q}}_c} \succeq {\bf{0}}\label{op4d},
\end{align}
\end{subequations}
in which $\tau ' \buildrel \Delta \over = {2^{\tau _{ms}}} - 1$, $\alpha $ is a slack variable introduced to simplify the denominator of the objective function in (\ref{op3}), and constraint (\ref{op4b}) is an equivalent form of (\ref{op3a}).

Next, we show that (\ref{op4}) can be recast as a two-stage optimization problem, and the outer problem is an one-variable optimization problem over $\alpha$. First, to achieve a non-negative secrecy rate, an upper bound of $\alpha$ can be determined via
\begin{equation}\label{upper2}
\alpha  \le 1 + \frac{{{{\bf{h}}_1}{{\bf{Q}}_c}{\bf{h}}_1^H}}{{1 + {{\bf{h}}_1}{{\bf{Q}}_a}{\bf{h}}_1^H}} \le 1 + {{\bf{h}}_1}{{\bf{Q}}_c}{\bf{h}}_1^H \le 1 + P{\left\| {{{\bf{h}}_1}} \right\|^2},
\end{equation}
where the third inequality follows from the fact that ${{\bf{h}}_1}{{\bf{Q}}_c}{\bf{h}}_1^H \le {\mathop{\rm Tr}} ({{\bf{Q}}_c}){\left\| {{{\bf{h}}_1}} \right\|^2}$ for any ${{\bf{Q}}_c} \succeq {\bf{0}}$ and ${\mathop{\rm Tr}} ({{\bf{Q}}_c}) \le P$. Since constraint (\ref{op4a}) can be expressed as
\begin{equation}\label{eq1}
  (\alpha  - 1)(1 + {{\bf{h}}_k}{{\bf{Q}}_a}{\bf{h}}_k^H) - {{\bf{h}}_k}{{\bf{Q}}_c}{\bf{h}}_k^H \ge 0,\forall k \in {{\cal K}_e}
\end{equation}
and $\log(\cdot )$ function is monotonically increasing, we further rewrite (\ref{op4}) as (\ref{op4.1}).
\begin{equation}\label{op4.1}
\begin{split}
\gamma ^*(\tau ') &= \mathop {\max }\limits_\alpha \;\eta (\tau ',\alpha )\\
\text{s.t.}\quad &1 \le \alpha  \le 1 + P{\left\| {{{\bf{h}}_1}} \right\|^2},
\end{split}
\end{equation}
where $\log {\gamma ^*}(\tau ') = {g^*}(\tau ') $, and
\begin{subequations}\label{op4.2}
\begin{align}
\nonumber&\eta (\tau ',\alpha ) = \mathop {\max}\limits_{{{\bf{Q}}_0},{{\bf{Q}}_a},{{\bf{Q}}_c}} \frac{{1 + {{\bf{h}}_1}({{\bf{Q}}_c} + {{\bf{Q}}_a}){\bf{h}}_1^H}}{{\alpha (1 + {{\bf{h}}_1}{{\bf{Q}}_a}{\bf{h}}_1^H)}}\\
\text{s.t.}\;&(\alpha  - 1)(1 + {{\bf{h}}_k}{{\bf{Q}}_a}{\bf{h}}_k^H) - {{\bf{h}}_k}{{\bf{Q}}_c}{\bf{h}}_k^H \ge 0,\forall k \in {{\cal K}_e},\label{op4.2a}\\
&{{\bf{h}}_k}{{\bf{Q}}_0}{\bf{h}}_k^H - \tau '{{\bf{h}}_k}({{\bf{Q}}_a} + {{\bf{Q}}_c}){\bf{h}}_k^H - \tau ' \ge 0,\forall k \in {\cal K},\label{op4.2b}\\
&\text{Tr}({{\bf{Q}}_0} + {{\bf{Q}}_a} + {{\bf{Q}}_c}) \le P,\label{op4.2c}\\
&{{\bf{Q}}_0} \succeq {\bf{0}}, {{\bf{Q}}_a} \succeq {\bf{0}}, {{\bf{Q}}_c} \succeq {\bf{0}}.\label{op4.2d}
\end{align}
\end{subequations}
We split (\ref{op4}) into two stages in (\ref{op4.1}) and (\ref{op4.2}): The maximization problem (\ref{op4.2}) is a quasiconvex problem, whose globally optimal solution can be searched by the bisection method \cite{boyd2009convex}. Even so, it is still preferred to solve (\ref{op4.2}) by reformulating it as a convex problem if possible. Fortunately, (\ref{op4.2}) indeed can be reformulated as a convex problem by applying the Charnes-Cooper transformation \cite{Charnes1962Programming}, i.e.,
\begin{equation}\label{cct}
{{\mathbf{Q}}_c} = {\mathbf{Z}}/\xi , {{\mathbf{Q}}_a} = {\mathbf{\Gamma }}/\xi , {{\mathbf{Q}}_0} = {\mathbf{\Phi }}/\xi , \xi > 0.
\end{equation}
Then we can rewrite (\ref{op4.2}) as an SDP problem, i.e.,
\begin{subequations}\label{op5}
\begin{align}
\nonumber\eta (\alpha ,\tau ') &= \mathop {\max}\limits_{{\bf{Z}},{\bf{\Gamma }},{\bf{\Phi }},\xi }\xi  + {{\bf{h}}_1}({\bf{Z}} + {\bf{\Gamma }}){\bf{h}}_1^H\\
\text{s.t.}\quad &\alpha\xi  + \alpha{{\bf{h}}_1}{\bf{\Gamma h}}_1^H = 1,\label{op5a}\\
&(\alpha  - 1)(\xi  + {{\bf{h}}_k}{\bf{\Gamma h}}_k^H) \ge {{\bf{h}}_k}{\bf{Zh}}_k^H,\forall k \in {{\cal K}_e},\label{op5b}\\
&{{\bf{h}}_k}{\bf{\Phi h}}_k^H - \tau '{{\bf{h}}_k}({\bf{\Gamma }} + {\bf{Z}}){\bf{h}}_k^H - \xi \tau ' \ge 0,\forall k \in {\cal K},\label{op5c}\\
&\text{Tr}({\bf{\Phi }} + {\bf{\Gamma }} + {\bf{Z}}) \le P\xi ,\label{op5d}\\
&{{\bf{Z}}} \succeq {\bf{0}}, {\bf{\Gamma }} \succeq {\bf{0}}, {\bf{\Phi }} \succeq {\bf{0}}\label{op5e}.
\end{align}
\end{subequations}
One can notice that the transformation turns (\ref{op4.2}) into a convex problem by fixing the denominator of $\eta (\tau ',\alpha )$. The convex problem (\ref{op5}) is an SDP problem, and thus can be efficiently solved through a convex optimization solver, e.g. \texttt{CVX} \cite{Boyd2011CVX}. Having obtained the optimal objective value for a fixed $\alpha$, the remnant work is simply adopting a proper one dimension search algorithm over $\alpha$. The golden section search \cite{Bertsekas1999} or uniform sampling search can be exploited to acquire the optimal $\alpha$ and ${\gamma ^*}(\tau ')$. The optimal $\alpha$ should be chosen as the one that leads to the maximum $\gamma ^*(\tau ')$ in (\ref{op4.1}). Ultimately, the optimal ${{\mathbf{Q}}_0}$, ${{\bf{Q}}_c}$ and ${{\bf{Q}}_a}$, denoted by $({\bf{Q}}_0^{*},{\bf{Q}}_c^{*},{\bf{Q}}_a^{*})$, can be retrieved through the relation (\ref{cct}).

\begin{remark}
Besides the aforementioned weighted sum method and our proposed QoMS-based method, some other scalarization methods have been proposed in literature to find the complete Pareto set for biobjective optimization, e.g., the weighted Tchebycheff method \cite{ng2016multi,marler2004survey}. However, this method would yield a nonconvex scalar optimization problem if used to tackle the specific scenario considered here, which is intractable or prohibitively time-consuming to solve. Therefore, this method may fail to reveal the complete Pareto optimal set.
\end{remark}

\subsection{Rank-Profile Analysis}
When the optimal solution $({\bf{Q}}_{0}^{*},{\bf{Q}}_{a}^{*},{\bf{Q}}_{c}^{*})$ to (\ref{op4.2}) satisfies the rank condition: $\text{rank}({\bf{Q}}_{0}^{*}) \le 1,\text{rank}({\bf{Q}}_{a}^{*}) \le 1$ and $\text{rank}({\bf{Q}}_{c}^{*}) \le 1$ for any given $\alpha$, the corresponding maximum secrecy rate $\gamma ^*(\tau ')$ could be attained via single-stream transmit beamforming, which facilitates the implementation of physically realizable transceiver with low complexity. Though the rank one properties cannot be generally fulfilled for ${\bf{Q}}_{0}^{*}$ and ${\bf{Q}}_{a}^{*}$, we give a proposition as below to guarantee $\text{rank}({\bf{Q}}_{c}^{*}) = 1$. Physically, it means that transmit beamforming is an optimal strategy for the transmission of confidential information.
\begin{proposition}\label{proposition1}
For problem (\ref{op4}), the optimal transmit covariance matrix of the confidential message, denoted by ${\mathbf{Q}}_c^{*}$, is rank-one.
\end{proposition}
\begin{IEEEproof}
The proof can be found in Appendix \ref{rank_proof1}.
\end{IEEEproof}

The exact investigation on rank properties of ${\bf{Q}}_{0}^{*}$ and ${\bf{Q}}_{a}^{*}$ still remains an open problem; thankfully, by employing some advanced results about SDP problems, we can prove that the rank one properties still hold for ${\bf{Q}}_0^{*}$ and ${\bf{Q}}_a^{*}$ in some special cases. Next a sufficient condition is given in the following proposition, under which ${\rm{rank}}({\bf{Q}}_0^{*}) = 1$ and ${\rm{rank}}({\bf{Q}}_a^{*}) \le 1$ will hold.
\begin{proposition}\label{proposition2}
If there only exists a single unauthorized receiver, i.e., $K-1=1$, then ${\rm{rank}}({\bf{Q}}_0^{*}) = 1,{\rm{rank}}({\bf{Q}}_a^{\rm{*}}) \le 1$.
\end{proposition}
\begin{IEEEproof}
In fact, Proposition \ref{proposition2} is an immediate result of \cite[Theorem 3.2]{Huang2010Rank}. The proof utilizes the solution equivalence of problems (\ref{op4.2}) and (\ref{opA1}). For (\ref{opA1}), it is a separable SDP problem \cite{Huang2010Rank}, thus satisfying
\begin{equation}\label{rank}
{\rm{ran}}{{\rm{k}}^2}({\bf{Q}}_0^{\rm{*}}){\rm{ + ran}}{{\rm{k}}^2}({\bf{Q}}_a^{\rm{*}}){\rm{ + ran}}{{\rm{k}}^2}({\bf{Q}}_c^{\rm{*}}) \le M,
\end{equation}
where $M$ denotes the total number of linear equalities and inequalities in (\ref{opA1}). For (\ref{opA1}), $M=2K$.

When $K=2$, incorporating ${\rm{rank}}({\bf{Q}}_c^{*}) = 1$, one can readily verify ${\rm{rank}}({\bf{Q}}_0^{\rm{*}}) \le 1,{\rm{rank}}({\bf{Q}}_a^{*}) \le 1$. Then we have completed the proof in that ${\bf{Q}}_0^{*} = {\bf{0}}$ is infeasible to (\ref{opA1}).
\end{IEEEproof}

\subsection{Complexity Analysis}
After giving the approach to finding the boundary points of the secrecy rate region (\ref{Region1}), we pay our attention to the complexity performance of our proposed method. Recall that for a given QoMS requirement, our proposed solution is derived from a two-stage optimization approach, the outer being one-dimensional search and the inner being SDP. The complexity of our proposed approach can be roughly calculated through the complexity of solving (\ref{op5}) times the number of searches involved, and times the number of boundary points we want to acquire. Let us take the uniform sampling search as an example, we characterize its maximum number of searches as follows.
\begin{proposition}\label{CA}
Let $\bar \alpha$ be an $\epsilon$-suboptimal solution of (\ref{op4.1}), satisfying ${g^*}(\tau ') - \log \eta (\tau ',\bar \alpha ) < \epsilon $, for some small positive constant $\epsilon$. If an uniform sampling search over $[1,1 + P{\left\| {{{\bf{h}}_1}} \right\|^2}]$ is exploited, one can find such $\bar \alpha$ with a maximum number of searches given by
\begin{equation}\label{SCtimes}
T_1 = \frac{{P{{\left\| {{{\bf{h}}_1}} \right\|}^2}}}{{{2^\varepsilon } - 1}}.
\end{equation}
Thus, the total arithmetic computation cost is on the order of
\[M_1=T_1\ln{(1/\epsilon)}\sqrt \gamma  \zeta,\]
where $\gamma$ and $\zeta$ are defined as below, and $n={\cal{O}}(3N_t^2+1)$.
\begin{align}\label{comp.order}
&\gamma = 3N_t + 2K + 1, \nonumber \\
&\zeta = n(3N_t^3 + 2K + 1)+ n^2(3N_t^2 + 2K + 1) + n^3
\end{align}
\end{proposition}
\begin{IEEEproof}
The proof can be found in Appendix \ref{CA_proof1}.
\end{IEEEproof}

To obtain $N$ boundary points of (\ref{Region1}), the total number of searches should be $M_N=NM_1$. Therefore, the total arithmetic computation cost of our proposed two-stage approach is polynomial w.r.t. the problem size for a given solution accuracy $\epsilon$.

\section{Extension: the Worst-case Robust SRRM}
Hitherto, we have assumed that the CSI can be perfectly obtained at the transmitter. We are now in a position to extend our model developed in the last section to an imperfect CSI case, where the transmitter has incomplete knowledge of all receivers' CSI. To capture the impact of the CSI imperfection and isolate specific channel estimation methods from the resource allocation algorithm design \cite{ng2016multi}, we consider a worst-case robust SRRM formulation under norm-bounded CSI uncertainties \cite{huang2012robust, yoo2006capacity} and develop an SDP-based optimization approach for the problem.

\subsection{The Worst-case Robust Problem Formulation}
We consider the same problem setup as in Section \Rmnum {2}, with a more general assumption that the transmitter has imperfect
CSI on links of all receivers. Let
\begin{equation}\label{err_csi}
{{\bf{h}}_k} = {{\bf{\tilde h}}_k} + {{\bf{e}}_k},{\left\| {{{\bf{e}}_k}} \right\|_F} \le {\varepsilon _k},\forall k \in {\cal K},
\end{equation}
where ${\bf h}_k$ is the actual channel vector between the transmitter and the $k$th receiver as defined before, ${{\bf{\tilde h}}_k}$ is the transmitter's estimation of ${\bf h}_k$, and ${{\bf{e}}_k}$ represents the associated CSI error which is located in a ball whose radius is ${\varepsilon _k}$. Here, we assume a nontrivial case where ${\varepsilon _k}$ is less than the norm of ${{\bf{\tilde h}}}_k$ for $\forall k \in {\cal K}$. The worst-case secrecy rate region is therefore determined by (cf. \cite{wyrembelski2014strong,liang2009information})
\begin{subequations}\label{WC_Region1}
\begin{align}
&{R_0} \le \mathop {\min }\limits_{k \in {\cal K}} {C_{m,k}^{\rm{worst}}}({{\bf{Q}}_0},{{\bf{Q}}_c},{{\bf{Q}}_a}),\\
&{R_c} \le {C_b^{\rm{worst}}}({{\bf{Q}}_c},{{\bf{Q}}_a})-\mathop {\max }\limits_{k \in {{\cal K}_e}} {C_{e,k}^{\rm{worst}}}({{\bf{Q}}_c},{{\bf{Q}}_a}),
\end{align}
\end{subequations}
where
\begin{subequations}\label{WC_Region2}
\begin{align}
&{C_{m,k}^{\rm{worst}}} \buildrel \Delta \over = \mathop {\min }\limits_{{{\bf{h}}_k} \in {B_k}} \log \left( {1 + \frac{{{{\bf{h}}_k}{{\bf{Q}}_0}{\bf{h}}_k^H}}{{1 + {{\bf{h}}_k}({{\bf{Q}}_c} + {{\bf{Q}}_a}){\bf{h}}_k^H}}} \right),\\
&{C_b^{\rm{worst}}} \buildrel \Delta \over = {\mathop {\min }\limits_{{{\bf{h}}_1} \in {B_1}} \log \left( {1 + \frac{{{{\bf{h}}_1}{{\bf{Q}}_c}{\bf{h}}_1^H}}{{1 + {{\bf{h}}_1}{{\bf{Q}}_a}{\bf{h}}_1^H}}} \right)},\\
&{C_{e,k}^{\rm{worst}}} \buildrel \Delta \over = {\mathop {\max }\limits_{{{\bf{h}}_k} \in {B_k}} \log \left( {1 + \frac{{{{\bf{h}}_k}{{\bf{Q}}_c}{\bf{h}}_k^H}}{{1 + {{\bf{h}}_k}{{\bf{Q}}_a}{\bf{h}}_k^H}}} \right)},
\end{align}
\end{subequations}
where ${B_k} \buildrel \Delta \over = \{ {{\bf{h}}_k}|{{{\bf{h}}_k} = {{{\bf{\tilde h}}}_k} + {{\bf{e}}_k},{{\left\| {{{\bf{e}}_k}} \right\|}_F} \le {\varepsilon _k}}\},\forall k \in {\cal K}$ denotes the set of all admissible CSIs. Physically, ${C_b^{\rm{worst}}}$ characterizes receiver 1's least possible mutual information among all admissible CSI in $B_1$, ${C_{e,k}^{\rm{worst}}}, k \in {{\cal K}_e}$ characterizes receiver $k$'s largest possible mutual information among all admissible CSI in $B_k$, and ${C_{m,k}^{\rm{worst}}}, k \in {\cal K}$ characterizes receiver $k$'s worst-case multicast rate among all admissible CSI in $B_k$. Therefore, the region (\ref{WC_Region1}) is a safe achievable region when the uncertainties given in (\ref{err_csi}) exists, and the actual secrecy rate pairs w.r.t. the true channel vectors must not lie within the boundary of (\ref{WC_Region1}).

Then, to obtain the robust design of $\mathbf{Q}_0$, $\mathbf{Q}_c$ and $\mathbf{Q}_a$, we focus on the following worst-case achievable SRRM problem,
\begin{subequations}\label{op6}
\begin{align}
\nonumber&\mathop {\max}\limits_{{{\bf{Q}}_0},{{\bf{Q}}_a},{{\bf{Q}}_c},R_0,R_c} \left({\text{w.r.t.}}\; {\mathbb {R}}_ + ^2 \right)\; \left( {{R_0},{R_c}} \right)\\
\text{s.t.}\quad &\mathop {\min }\limits_{k \in {\cal K}} {C_{m,k}^{\rm{worst}}}({{\bf{Q}}_0},{{\bf{Q}}_c},{{\bf{Q}}_a}) \ge {R_0} ,\label{op1a}\\
& {C_b^{\rm{worst}}}({{\bf{Q}}_c},{{\bf{Q}}_a})-\mathop {\max }\limits_{k \in {{\cal K}_e}} {C_{e,k}^{\rm{worst}}}({{\bf{Q}}_c},{{\bf{Q}}_a}) \ge {R_c},\\
&\text{Tr}({{\bf{Q}}_0} + {{\bf{Q}}_a} + {{\bf{Q}}_c}) \le P,\\
&{{\bf{Q}}_0} \succeq {\bf{0}}, {{\bf{Q}}_a} \succeq {\bf{0}}, {{\bf{Q}}_c} \succeq {\bf{0}}.
\end{align}
\end{subequations}
One can check that plunging (\ref{WC_Region2}) into (\ref{op6}) yields
\begin{subequations}\label{op7}
\begin{align}
\nonumber&\mathop {\max}\limits_{{{\bf{Q}}_0},{{\bf{Q}}_a},{{\bf{Q}}_c},R_0,R_c} \left({\text{w.r.t.}}\; {\mathbb {R}}_ + ^2 \right)\; \left( {{R_0},{R_c}} \right)\\
\text{s.t.}\quad &\mathop {\min}\limits_{k \in {\cal K}} \mathop {\min}\limits_{{{\bf{h}}_k} \in {B_k}}\;\log \left( {1 + \frac{{{{\bf{h}}_k}{{\bf{Q}}_0}{\bf{h}}_k^H}}{{1 + {{\bf{h}}_k}({{\bf{Q}}_c} + {{\bf{Q}}_a}){\bf{h}}_k^H}}} \right) \ge {R_0}, \label{op7a}\\
&\nonumber \mathop {\min }\limits_{{{\bf{h}}_1} \in {B_1}} \log \left( {1 + \frac{{{{\bf{h}}_1}{{\bf{Q}}_c}{\bf{h}}_1^H}}{{1 + {{\bf{h}}_1}{{\bf{Q}}_a}{\bf{h}}_1^H}}} \right) - \\
&\quad \mathop {\max }\limits_{{{\bf{h}}_k} \in {B_k}} \log \left( {1 + \frac{{{{\bf{h}}_k}{{\bf{Q}}_c}{\bf{h}}_k^H}}{{1 + {{\bf{h}}_k}{{\bf{Q}}_a}{\bf{h}}_k^H}}} \right) \ge {R_c}, \forall k \in {{\cal K}_e}\\
&\text{Tr}({{\bf{Q}}_0} + {{\bf{Q}}_a} + {{\bf{Q}}_c}) \le P,\\
&{{\bf{Q}}_0} \succeq {\bf{0}}, {{\bf{Q}}_a} \succeq {\bf{0}}, {{\bf{Q}}_c} \succeq {\bf{0}}.
\end{align}
\end{subequations}
Due to the existence of uncertainties in the constraints, the vector optimization problem (\ref{op7}) appears more intricate to solve than (\ref{op2}). As a routine, we degrade (\ref{op7}) into a standard scalar optimization problem using the same procedures we adopted in Section \Rmnum{3}.

\subsection{An Equivalent Scalar Optimization Problem of (\ref{op7})}
Similar to Section \Rmnum {3}.A, we first fix the variable $R_0$ as a constant $\tau _{ms} \ge 0$. As a result, the degraded version of (\ref{op7}) is given as below.
\begin{subequations}\label{op8}
\begin{align}
\nonumber \mathop {\max}\limits_{{{\bf{Q}}_0},{{\bf{Q}}_a},{{\bf{Q}}_c}} & \log \frac{{\mathop {\min }\limits_{{{\bf{h}}_1} \in {B_1}} 1 + {{(1 + {{\bf{h}}_1}{{\bf{Q}}_a}{\bf{h}}_1^H)}^{ - 1}}{{\bf{h}}_1}{{\bf{Q}}_c}{\bf{h}}_1^H}}{{\mathop {\max }\limits_{k \in {{\cal K}_e}, {{\bf{h}}_k} \in {B_k}} 1 + {{(1 + {{\bf{h}}_k}{{\bf{Q}}_a}{\bf{h}}_k^H)}^{ - 1}}{{\bf{h}}_k}{{\bf{Q}}_c}{\bf{h}}_k^H}}\\
\text{s.t.}\; &\mathop {\min}\limits_{k \in {\cal K}} \mathop {\min}\limits_{{{\bf{h}}_k} \in {B_k}}\;\log \left( {1 + \frac{{{{\bf{h}}_k}{{\bf{Q}}_0}{\bf{h}}_k^H}}{{1 + {{\bf{h}}_k}({{\bf{Q}}_c} + {{\bf{Q}}_a}){\bf{h}}_k^H}}} \right) \ge {\tau _{ms}}, \label{op8a}\\
&\text{Tr}({{\bf{Q}}_0} + {{\bf{Q}}_a} + {{\bf{Q}}_c}) \le P,\\
&{{\bf{Q}}_0} \succeq {\bf{0}}, {{\bf{Q}}_a} \succeq {\bf{0}}, {{\bf{Q}}_c} \succeq {\bf{0}},
\end{align}
\end{subequations}
where the variable $R_c$ is discarded as a slack variable again. We also gain some insights on the formulation of (\ref{op8}): $\tau _{ms}$ is preset requirement of the least achievable multicast rate, and (\ref{op8}) is a worst-case robust SRM problem with worst-case QoMS constraints. By setting $\tau _{ms}=0$, (\ref{op8}) becomes a conventional AN-aided worst-case robust SRM problem for multi-user MISO system. The maximum value of $\tau _{ms}$, denoted by $\tau^{\rm{worst}}_{\max}$, is attained when the confidential message transmission is terminated, i.e.,
\begin{equation}\label{max_tau2}
{\tau^{\rm{worst}}_{\max }} = \mathop {\max }\limits_{{{\bf{Q}}_0} \succeq {\bf{0}}, \text{Tr}({{\bf{Q}}_0}) \le P} \mathop {\min }\limits_{k \in {\cal {K}}, {{\bf{h}}_k} \in {B_k}} \log (1 + {{\bf{h}}_k}{{\bf{Q}}_0}{\bf{h}}_k^H),
\end{equation}
where $\tau^{\rm{worst}}_{\max}$ is essentially the largest achievable worst-case multicast rate. The optimization problem (\ref{max_tau}) can also be solved via an SDP reformulation; see, e.g., \cite{shenouda2007convex}.

One can notice that the maximum and minimum in the objective function of (\ref{op8}) have no effect on the efficacy of our construction method adopted in the proof of Theorem \ref{theorem1}. Therefore, by reusing the procedures we introduce in the proof of Theorem \ref{theorem1}, it is straightforward for us to obtain the following properties w.r.t. (\ref{op8}) and Theorem \ref{theorem2}.
\begin{property}\label{property3}
The maximum objective value of problem (\ref{op8}) is obtained only when the equality in (\ref{op8a}) holds.
\end{property}
\begin{property}\label{property4}
The optimal objective value of (\ref{op8}), denoted as ${g^*}({\tau _{ms}})$, is monotonically decreasing w.r.t. $\tau _{ms}$.
\end{property}

\begin{theorem}\label{theorem2}
The rate pair $({\tau _{ms}},{g^*}({\tau _{ms}}))$ is a Pareto optimal point of (\ref{op7}), and all Pareto optimal points of (\ref{op7}) can be obtained by varying $\tau _{ms}$'s lying within $[0,\tau^{\rm{worst}}_{\max}]$.
\end{theorem}
\begin{figure*}[!t]
\normalsize
\setcounter{MYtempeqncnt}{\value{equation}}
\setcounter{equation}{26}
\begin{equation}\label{lmi1}
{{\bf{T}}_k}(\beta ,{{\bf{Q}}_c},{{\bf{Q}}_a},{t_k}) = \left[ {\begin{array}{*{20}{c}}
{{t_k}{\bf{I}} + (\beta  - 1){{\bf{Q}}_a} - {{\bf{Q}}_c}}&{((\beta  - 1){{\bf{Q}}_a} - {{\bf{Q}}_c}){\bf{\tilde h}}_k^H}\\
{{{{\bf{\tilde h}}}_k}((\beta  - 1){{\bf{Q}}_a} - {{\bf{Q}}_c})}&{{{{\bf{\tilde h}}}_k}((\beta  - 1){{\bf{Q}}_a} - {{\bf{Q}}_c}){\bf{\tilde h}}_k^H - {t_k}\varepsilon _k^2 + \beta  - 1}
\end{array}} \right]\succeq {\bf{0}},\forall k \in {{\cal K}_e},
\end{equation}
\begin{equation}\label{lmi2}
{{\bf{S}}_k}({{\bf{Q}}_c},{{\bf{Q}}_a},{{\bf{Q}}_0},{\delta _k}) = \left[ {\begin{array}{*{20}{c}}
{{\delta _k}{\bf{I}} + {{\bf{Q}}_0} - \tau '({{\bf{Q}}_a} + {{\bf{Q}}_c})}&{({{\bf{Q}}_0} - \tau '({{\bf{Q}}_a} + {{\bf{Q}}_c})){\bf{\tilde h}}_k^H}\\
{{{{\bf{\tilde h}}}_k}({{\bf{Q}}_0} - \tau '({{\bf{Q}}_a} + {{\bf{Q}}_c}))}&{ - {\delta _k}\varepsilon _k^2 - \tau ' + {{{\bf{\tilde h}}}_k}({{\bf{Q}}_0} - \tau '({{\bf{Q}}_a} + {{\bf{Q}}_c})){\bf{\tilde h}}_k^H}
\end{array}} \right] \succeq {\bf{0}},\forall k \in {\cal K}.
\end{equation}
\hrulefill
\setcounter{equation}{\value{MYtempeqncnt}}
\vspace*{-12pt}
\end{figure*}
\begin{figure*}[!b]
\normalsize
\setcounter{MYtempeqncnt}{\value{equation}}
\setcounter{equation}{32}
\vspace*{-5pt}
\hrulefill
\vspace*{12pt}
\begin{equation}\label{lmi3}
{\bf{U}}(\beta ,{{\bf{Q}}_c},{{\bf{Q}}_a},\rho ) = \left[ {\begin{array}{*{20}{c}}
{\rho{\bf{I}} + {{\bf{Q}}_c} + (1-\alpha\beta){{\bf{Q}}_a}}&{({{\bf{Q}}_c} + (1-\alpha\beta){{\bf{Q}}_a}){\bf{\tilde h}}_k^H}\\
{{{{\bf{\tilde h}}}_k}({{\bf{Q}}_c} + (1-\alpha\beta){{\bf{Q}}_a})}&{{{{\bf{\tilde h}}}_k}({{\bf{Q}}_c} + (1-\alpha\beta){{\bf{Q}}_a}){\bf{\tilde h}}_k^H - \rho\varepsilon _1^2 - \alpha\beta  + 1}
\end{array}} \right]\succeq {\bf{0}}.
\end{equation}
\setcounter{equation}{\value{MYtempeqncnt}}
\vspace*{-12pt}
\end{figure*}

\subsection{A Tractable Reformulation of (\ref{op8})}
Our next endeavor is to develop a tractable reformulation of (\ref{op8}) that reveals its hidden convexity and thus caters to the numerical optimization. To start with, by introducing the slack variables $\beta$, we rewrite (\ref{op8}) as
\begin{subequations}\label{op9}
\begin{align}
\nonumber{g^*}(\tau ') &= \mathop {\max}\limits_{{{\bf{Q}}_0},{{\bf{Q}}_a},{{\bf{Q}}_c},\beta} \mathop {\min }\limits_{{{\bf{h}}_1} \in {B_1}} \log\left( \frac{{1 + {{\bf{h}}_1}({{\bf{Q}}_c} + {{\bf{Q}}_a}){\bf{h}}_1^H}}{{\beta (1 + {{\bf{h}}_1}{{\bf{Q}}_a}{\bf{h}}_1^H)}} \right)\\
\text{s.t.}\; &\log \left(1 + \frac{{{{\bf{h}}_k}{{\bf{Q}}_c}{\bf{h}}_k^H}}{{1 + {{\bf{h}}_k}{{\bf{Q}}_a}{\bf{h}}_k^H}}\right) \le \log\beta ,\forall k \in {{\cal K}_e},{{\bf{h}}_k} \in {B_k},\label{op9c}\\
&\frac{{{{\bf{h}}_k}{{\bf{Q}}_0}{\bf{h}}_k^H}}{{1 + {{\bf{h}}_k}({{\bf{Q}}_c} + {{\bf{Q}}_a}){\bf{h}}_k^H}} \ge \tau ',\forall k \in {\cal K},{{\bf{h}}_k} \in {B_k},\label{op9d}\\
&{\rm{Tr}}({{\bf{Q}}_0} + {{\bf{Q}}_a} + {{\bf{Q}}_c}) \le P,\\
&{{\bf{Q}}_0} \succeq {\bf{0}},{{\bf{Q}}_a} \succeq {\bf{0}},{{\bf{Q}}_c} \succeq {\bf{0}},
\end{align}
\end{subequations}
in which $\beta \ge 1$, $\tau ' \buildrel \Delta \over = {2^{\tau_{ms}} } - 1$, and thus constraint (\ref{op9d}) is an equivalent form of constraint (\ref{op8a}). One can notice that $\beta $ is introduced to simplify the denominator of the logarithm in the objective function of (\ref{op1}). Currently, the obstacle of dealing with (\ref{op9}) lies in the existence of uncertainties in the objective function and the constraints (\ref{op9c}) and (\ref{op9d}). To deal with the uncertainties, we first exert ${\cal S}$-procedure \cite{boyd2009convex} to turn the constraints (\ref{op9c}) and (\ref{op9d}) into linear matrix inequalities (LMIs) in (\ref{lmi1}) and \addtocounter{equation}{2}(\ref{lmi2}) at the top of this page, where ${\left\{ {{t_k}} \right\}_{k \in {{\cal K}_e}}}$ and ${\left\{ {{\delta _k}} \right\}_{k \in {\cal K}}}$ are all nonnegative slack variables.

Next, we show that (\ref{op9}) can be recast as a one-variable optimization problem over $\beta$ which involves solving a quasiconcave problem. Analogous to (\ref{upper2}), to achieve a non-negative secrecy rate, an upper bound on $\beta$ can be determined via
\begin{equation}\label{beta_upper}
\begin{split}
\beta  &\le 1 + \mathop {\min }\limits_{{{\bf{h}}_1} \in {B_1}} \frac{{{{\bf{h}}_1}{{\bf{Q}}_c}{\bf{h}}_1^H}}{{1 + {{\bf{h}}_1}{{\bf{Q}}_a}{\bf{h}}_1^H}} \\
&\le 1 + \mathop {\min }\limits_{{{\bf{h}}_1} \in {B_1}} {{\bf{h}}_1}{{\bf{Q}}_c}{\bf{h}}_1^H \le 1 + P\mathop {\min }\limits_{{{\bf{h}}_1} \in {B_1}} {\left\| {{{\bf{h}}_1}} \right\|^2}\\
&= 1 + P({\lVert {{{{\bf{\tilde h}}}_1}} \rVert-{\varepsilon _1})^2},
\end{split}
\end{equation}
where the last equality is derived by solving a simple quadratically constrained quadratic programming (QCQP) with its Karush-Kuhn-Tucker (KKT) conditions, which leads to one upper bound on $\beta$.

Noting that $\log(\cdot )$ function is monotonically increasing, we further rewrite (\ref{op9}) as
\begin{equation}\label{op9.1}
\begin{split}
\gamma ^*(\tau ') &= \mathop {\max }\limits_\beta \;\eta (\tau ',\beta )\\
\text{s.t.}\quad &1 \le \beta  \le \beta_{\max},
\end{split}
\end{equation}
where $\log {\gamma ^*}(\tau ') = {g^*}(\tau ') $, $\beta_{\max} \buildrel \Delta \over = 1 + P({\lVert {{{{\bf{\tilde h}}}_1}} \rVert-{\varepsilon _1})^2}$, and
\begin{subequations}\label{op9.2}
\begin{align}
\nonumber\eta (\tau ',\beta ) &= \mathop {\max }\limits_{{{\bf{Q}}_0},{{\bf{Q}}_a},{{\bf{Q}}_c}\atop{\left\{ {{t_k}} \right\}_{k \in {{\cal K}_e}}},{\left\{ {{\delta _k}} \right\}_{k \in {\cal K}}}} \mathop {\min }\limits_{{{\bf{h}}_1} \in {B_1}} \frac{{1 + {{\bf{h}}_1}({{\bf{Q}}_c} + {{\bf{Q}}_a}){\bf{h}}_1^H}}{{\beta (1 + {{\bf{h}}_1}{{\bf{Q}}_a}{\bf{h}}_1^H)}}\\
\text{s.t.}\quad &{{\bf{T}}_k}(\beta ,{{\bf{Q}}_c},{{\bf{Q}}_a},{t_k}) \succeq {\bf{0}},{t_k} \ge 0,\forall k \in {{\cal K}_e},\label{op9.2a}\\
&{{\bf{S}}_k}({{\bf{Q}}_c},{{\bf{Q}}_a},{{\bf{Q}}_0},{\delta _k}) \succeq {\bf{0}},{\delta _k} \ge 0,\forall k \in {\cal K},\label{op9.2b}\\
&{\rm{Tr}}({{\bf{Q}}_0} + {{\bf{Q}}_a} + {{\bf{Q}}_c}) \le P,\label{op9.2c}\\
&{{\bf{Q}}_0} \succeq {\bf{0}},{{\bf{Q}}_a} \succeq {\bf{0}},{{\bf{Q}}_c} \succeq {\bf{0}}\label{op9.2d}.
\end{align}
\end{subequations}

To proceed, we will next show the maximization problem (\ref{op9.2}) is a quasiconcave maximization problem; thus, its global optimum can be efficiently found by using the bisection method \cite{boyd2009convex}. For ease of exposition, we first define
\[f({{{\bf{Q}}_a},{{\bf{Q}}_c}})=\mathop {\min }\limits_{{{\bf{h}}_1} \in {B_1}} \frac{{1 + {{\bf{h}}_1}({{\bf{Q}}_c} + {{\bf{Q}}_a}){\bf{h}}_1^H}}{{\beta (1 + {{\bf{h}}_1}{{\bf{Q}}_a}{\bf{h}}_1^H)}}.\]
With a slight abuse of notations but for notational simplicity, we replace $f({{{\bf{Q}}_a},{{\bf{Q}}_c}})$ by $f$ in the following section.
\begin{property}\label{property5}
$f$ is a quasiconcave function on the problem domain of (\ref{op9.2}), and hence the maximization problem (\ref{op9.2}) is a quasiconcave problem.
\end{property}
\begin{IEEEproof}
With the problem domain of (\ref{op9.2}) being convex, to verify Property \ref{property5}, we should check whether all the $\alpha$-superlevel sets of $f$ are convex for every $\alpha$ \cite{boyd2009convex}. The $\alpha$-superlevel set of $f$ is defined as
\begin{equation}\label{super}
{{\cal {S}}_{\alpha}}=\left\{ {\left( {{{\bf{Q}}_a},{{\bf{Q}}_c}} \right)\left| {{{\bf{Q}}_a} \succeq {\bf{0}},{{\bf{Q}}_c} \succeq {\bf{0}},} \right.f \ge \alpha } \right\}.
\end{equation}
Again, we resort to the $\cal S$-procedure for revealing the hidden convexity of the function $f \ge \alpha $, which is shown in \addtocounter{equation}{1}(\ref{lmi3}) at the bottom of this page, in which $\rho$ is a slack variable satisfying $\rho \ge 0$. Equation (\ref{lmi3}) is an LMI, and convex to $\left({{\bf{Q}}_a}, {{\bf{Q}}_c}, \rho\right)$. Hence, ${{\cal {S}}_{\alpha}}$ is a convex set for every $\alpha$, and we know $f$ is a quasiconcave function, which completes our proof.
\end{IEEEproof}

Summarizing our reformulation of (\ref{op9}), we split (\ref{op9}) into two stages in (\ref{op9.1}) and (\ref{op9.2}): The maximization problem (\ref{op9.2}) is a quasiconcave problem and calculates $\eta (\tau ',\beta )$ for a fixed $\beta$, which can be efficiently solved by combining the bisection method with the convex optimization solver \texttt{CVX}. Its searching lower bound and upper bound can be chosen as $1/\beta$ and $\beta_{\max}/\beta$, respectively (cf. (\ref{beta_upper})). The outer problem (\ref{op9.1}) is a single-variable optimization problem with a bounded interval constraint $[1,\beta_{\max}]$, which can be handled by performing a proper one-dimensional search algorithm, and the procedure is the same as that described in Section \Rmnum{3}-B.

\subsection{Rank-Profile Analysis}
We now pay our attention to the rank properties of the optimal solution $({\bf{Q}}_{0}^{*},{\bf{Q}}_{a}^{*},{\bf{Q}}_{c}^{*})$ of problem (\ref{op9.2}). Particularly, one may curious about whether the rank-one property of ${\bf{Q}}_{c}^{*}$ applies to the imperfect CSI case. This issue could be solved in the following proposition.
\begin{proposition}\label{proposition4}
With AN and imperfect CSI on all links, the optimal transmit covariance matrix of the confidential message is still of rank one.
\end{proposition}
\begin{IEEEproof}
The proof can be found in Appendix \ref{rank_proof2}.
\end{IEEEproof}
\begin{table*}[t]
\centering
\caption{Computational Complexity of Proposed Schemes}\label{comp.all}
\begin{tabular}{|c|c|}
\hline
Scheme & Complexity Order (suppressing $\ln{(1/\epsilon)}$) \\
\hline
$\begin{array}{c}\text{Optimal scheme}\\ \text{(perfect CSI)}\end{array}$ & ${\cal{O}}\left(T_1\sqrt{3N_t + 2K + 1}[n(3N_t^3 + 2K + 1)+ n^2(3N_t^2 + 2K + 1) + n^3]\right)$, where $n={\cal{O}}(3N_t^2+1)$. \\
\hline
$\begin{array}{c}\text{Power splitting scheme}\\ \text{(perfect CSI)}\end{array}$ & ${\cal{O}}\left(T_1\sqrt{2N_t + K + 1}[n(2N_t^3 + K + 1)+ n^2(2N_t^2 + K + 1) + n^3]\right)$, where $n={\cal{O}}(2N_t^2+1)$. \\
\hline
$\begin{array}{c}\text{Optimal scheme}\\ \text{(imperfect CSI)}\end{array}$ & $\begin{array}{l}
{\cal{O}}\left( {T_1^{{\rm{wc}}}\sqrt {(2K + 3){N_t} + 4K} [{n^3} + {n^2}(2K{{({N_t} + 1)}^2} + 3N_t^2 + 2K) + n(2K{{({N_t} + 1)}^3} + 3N_t^3 + 2K)]} \right),\\
\text{where}\; n = {\cal{O}}(3N_t^2 + 2K - 1).
\end{array}$ \\
\hline
$\begin{array}{c}\text{Power splitting scheme}\\ \text{(imperfect CSI)}\end{array}$ & $\begin{array}{l}
{\cal{O}}\left( {T_1^{{\rm{wc}}}\sqrt {(K + 2){N_t} + 2K} [{n^3} + {n^2}(K{{({N_t} + 1)}^2} + 2N_t^2 + K) + n(K{{({N_t} + 1)}^3} + 2N_t^3 + K)]} \right),\\
\text{where}\; n = {\cal{O}}(2N_t^2 + K - 1).
\end{array}$ \\
\hline
$\begin{array}{c}\text{Lower bound based scheme}\\ \text{(imperfect CSI)}\end{array}$ & $\begin{array}{l}
{\cal{O}}\left( {T_1^{{\rm{lb}}}\sqrt {(2K + 4){N_t} + 4K + 3} [{n^3} + {n^2}((2K+1){{({N_t} + 1)}^2} + 3N_t^2 + 2K+2) + n((2K+1){{({N_t} + 1)}^3} } \right.\\
\left. {+ 3N_t^3+ 2K + 2)]} \right),\text{where}\; n = {\cal{O}}(3N_t^2 + 2K + 3)\; \text{and}\; T_1^{{\rm{lb}}}=\frac{P({\lVert {{{{\bf{\tilde h}}}_1}} \rVert-{\varepsilon _1})^2}}{{{2^\epsilon } - 1}}.
\end{array}$ \\
\hline
\end{tabular}
\end{table*}

\subsection{Complexity Analysis}
The process of characterizing the maximum number of searches for the imperfect CSI case is practically analogous to that in the perfect case. However, since the bisection method is adopted to find $\eta (\tau ',\beta )$, it will increase the total searching times. Another consideration is that the bisection method would introduce inaccuracy of $\eta (\tau ',\beta )$, relying on the preset convergence tolerance. If such convergence tolerance is set sufficiently loose, we may not guarantee the existence of an $\epsilon$-suboptimal solution for any $\epsilon > 0$. Still we take the uniform sampling search as an example, we characterize its maximum number of searches as follows in Proposition \ref{WCCA}.
\begin{proposition}\label{WCCA}
Let $\bar \beta$ be an $\epsilon$-suboptimal solution of (\ref{op9.1}), satisfying ${g^*}(\tau ') - \log \eta (\tau ',\bar \beta ) < \epsilon $, for some small positive constant $\epsilon$. If we exploit an uniform sampling search over $[1,\beta_{\max}]$ in (\ref{op9.1}) and a bisection method over $[1/\beta, \beta_{\max}/\beta]$ in (\ref{op9.2}), with the convergence tolerance of the bisection method set as $\epsilon_b$, then one can find such $\bar \beta$ with a maximum number of searches given by
\begin{equation}\label{SCtimes}
T_1^{\rm{wc}} = \sum\limits_{i = 1}^{{M_u}} {\log \left( {\frac{P({\lVert {{{{\bf{\tilde h}}}_1}} \rVert-{\varepsilon _1})^2}}{{(1 + \Delta i){\epsilon _b}}}} \right)},
\end{equation}
where \[M_u = \frac{(1 + {2^\epsilon }{\epsilon _b})P({\lVert {{{{\bf{\tilde h}}}_1}} \rVert-{\varepsilon _1})^2}}{{{2^\varepsilon }(1 - {\epsilon _b}) - 1}}, \Delta  = \frac{{{2^\epsilon }(1 - {\epsilon _b}) - 1}}{{1 + {2^\epsilon }{\epsilon _b}}}.\]
Thus, the total arithmetic computation cost is on the order of
\[M_1^{\rm{wc}}=T_1^{\rm{wc}}\ln{(1/\epsilon)}\sqrt \gamma  \zeta,\]
where $\gamma$ and $\zeta$ are defined as below, and $n={\cal{O}}(3N_t^2+2K-1)$.
\begin{equation}\label{wc.order}
\begin{split}
&\gamma = (2K + 3)N_t + 4K, \\
&\zeta = n^3+n^2(2K{({N_t} + 1)^2} + 3N_t^2 + 2K)\\
&\quad+n(2K{({N_t} + 1)^3} + 3N_t^3 + 2K)
\end{split}
\end{equation}
\end{proposition}
\begin{IEEEproof}
The proof can be found in Appendix \ref{CA_proof2}.
\end{IEEEproof}

One can notice from Proposition \ref{WCCA} that to achieve the $\epsilon$-suboptimality, the convergence tolerance of the bisection method must satisfy $\Delta > 0$, or equivalently, ${\epsilon _b} < 1 - {2^{ - \epsilon }}$.

Obviously, if we want to obtain $N$ boundary points of (\ref{WC_Region1}), the total number of searches should amount to $M_N^{\rm{wc}}=NM_1^{\rm{wc}}$. Then we know that the total arithmetic computation cost of our proposed two-stage approach, for the imperfect CSI case, is still polynomial w.r.t. the problem size for a given solution accuracy $\epsilon$.

\section{Suboptimal Schemes and Extensions}
In this section, we propose two suboptimal resource allocation schemes to implement PHY-SI in a more efficient manner. Then we briefly discuss two possible extensions of the methods introduced in the preceding sections.

\subsection{Power Splitting Scheme}
Our first proposed suboptimal scheme aims to decouple the multicast message transmission and the confidential message transmission by introducing a power splitting factor $\rho$ ($0 \le \rho \le 1$), such that $\text{Tr}({{\bf{Q}}_c}+{{\bf{Q}}_a})=\rho P$ and $\text{Tr}({{\bf{Q}}_0})=(1-\rho)P$. Then we specify a secrecy rate $R_c(\rho)$ using the power allocated to the confidential message and AN, and find the maximum multicast rate $R_0(\rho)$ the remaining transmit power can achieve. In the following, we take the imperfect CSI case as an example to show how to implement this scheme.

Specifically, $R_c(\rho)$ is chosen as the maximum worst-case secrecy rate with $\text{Tr}({{\bf{Q}}_c}+{{\bf{Q}}_a})=\rho P$. This worst-case SRM problem has been previously tackled in \cite{li2013spatially}. Then, let us denote the corresponding optimal ${{\bf{Q}}_c}$ and ${{\bf{Q}}_a}$ as ${{\bf{Q}}_c(\rho)}$ and ${{\bf{Q}}_a(\rho)}$, respectively. Next we will determine the maximum worst-case multicast rate with $\text{Tr}({{\bf{Q}}_0})=(1-\rho) P$, which can be obtained by solving the following optimization problem,
\begin{equation}\label{PS.multicast}
{\eta_0(\rho)} = \mathop {\max }\limits_{\text{Tr}({{\bf{Q}}_0}) \le (1-\rho)P \atop {{\bf{Q}}_0} \succeq {\bf{0}}} \mathop {\min }\limits_{k \in {\cal {K}}, {{\bf{h}}_k} \in {B_k}} \frac{{{\bf{h}}_k}{{\bf{Q}}_0}{\bf{h}}_k^H}{1+{{\bf{h}}_k}({{\bf{Q}}_a(\rho)}+{{\bf{Q}}_c(\rho)}){\bf{h}}_k^H},
\end{equation}
with $R_0(\rho)=\log(1+{\eta_0(\rho)})$. Problem (\ref{PS.multicast}) is a convex optimization problem after reformulating it as its epigraph form and reapplying the $\cal{S}$-procedure. Finally, traversing all $\rho$ lying within the interval $[0,1]$ will give rise to the secrecy rate region achieved by this power splitting scheme.


\subsection{A Computationally Efficient Lower Bound for the Worst-Case SRRM}
The purpose of the second suboptimal scheme is to reduce the computational complexity in solving the worst-case SRRM problem. As we can see from Proposition \ref{WCCA}, solving the worst-case SRRM problem involves a two-dimensional search, which renders the proposed methods time-consuming. Noting the following relation, i.e.,
\begin{align}\label{lower.bound}
&f({{{\bf{Q}}_a},{{\bf{Q}}_c}})=\mathop {\min }\limits_{{{\bf{h}}_1} \in {B_1}} \frac{{1 + {{\bf{h}}_1}({{\bf{Q}}_c} + {{\bf{Q}}_a}){\bf{h}}_1^H}}{{\beta (1 + {{\bf{h}}_1}{{\bf{Q}}_a}{\bf{h}}_1^H)}}\nonumber\\
&\ge \frac{{1 + \mathop {\min }\limits_{{{\bf{h}}_1} \in {B_1}} {{\bf{h}}_1}({{\bf{Q}}_c} + {{\bf{Q}}_a}){\bf{h}}_1^H}}{{\beta (1 + \mathop {\max }\limits_{{{\bf{h}}_1} \in {B_1}} {{\bf{h}}_1}{{\bf{Q}}_a}{\bf{h}}_1^H)}}\buildrel \Delta \over = \tilde f({{{\bf{Q}}_a},{{\bf{Q}}_c}}),
\end{align}
we propose to maximize $\tilde f({{{\bf{Q}}_a},{{\bf{Q}}_c}})$ in (\ref{op9.2}) to find a lower bound on $\eta (\tau ',\beta )$. The maximization of $\tilde f({{{\bf{Q}}_a},{{\bf{Q}}_c}})$ can be further reformulated into a convex optimization problem. To elaborate a little further, we can introduce two slack variables $u$ and $v$ to simplify the numerator and denominator of $\tilde f({{{\bf{Q}}_a},{{\bf{Q}}_c}})$ and rewrite (\ref{op9.2}) as
\begin{subequations}\label{op9.3}
\begin{align}
\nonumber \mathop {\max }\limits_{{{\bf{Q}}_0},{{\bf{Q}}_a},{{\bf{Q}}_c},u,v\atop{\left\{ {{t_k}} \right\}_{k \in {{\cal K}_e}}},{\left\{ {{\delta _k}} \right\}_{k \in {\cal K}}}}&uv^{-1}\\
\text{s.t.}\quad &{1 + \mathop {\min }\limits_{{{\bf{h}}_1} \in {B_1}} {{\bf{h}}_1}({{\bf{Q}}_c} + {{\bf{Q}}_a}){\bf{h}}_1^H} \ge u, \label{op9.3a}\\
&{\beta (1 + \mathop {\max }\limits_{{{\bf{h}}_1} \in {B_1}} {{\bf{h}}_1}{{\bf{Q}}_a}{\bf{h}}_1^H)} \le v, \label{op9.3b}\\
&\text{(\ref{op9.2a})-(\ref{op9.2d}) satisfied.}
\end{align}
\end{subequations}
To proceed, we introduce the following variable transformation, i.e.,
\begin{align}
&\xi=1/v, a=u/v, {{\mathbf{Q}}_c} = {\mathbf{Z}}/\xi , {{\mathbf{Q}}_a} = {\mathbf{\Gamma }}/\xi , {{\mathbf{Q}}_0} = {\mathbf{\Phi }}/\xi ,\nonumber\\
&{t_k} = {\lambda _k}/\xi ,\forall k \in {{\cal K}_e}, {\delta_k} = {\mu _k}/\xi ,\forall k \in {{\cal K}}.
\end{align}
Then one can verify that problem (\ref{op9.3}) can be recast as a convex problem after carrying out the transformation above. It is evident to see this suboptimal scheme is significantly more time efficient than the optimal one proposed in the last section. As an additional merit, this scheme may be asymptotically optimal at high QoMS region, since AN gradually diminishes with the increase in QoMS.

For ease of comparison, we summarize the computational complexity of our proposed optimal and suboptimal schemes in Table \ref{comp.all}, shown at the top of last page. Since in the power splitting scheme, the computation of $R_c(\rho)$ requires higher complexity than that of $R_0(\rho)$, the power splitting scheme should possess the same complexity order as computing $R_c(\rho)$. In Table \ref{comp.all}, the complexity order of maximizing the lower bound (\ref{lower.bound}) is derived by following the similar procedures to the proof of Proposition \ref{CA}, but the details are omitted here due to the page limit. One can see from Table \ref{comp.all} that the above-developed suboptimal schemes are more time-efficient to implement than the optimal ones.

\subsection{Extensions}
For simplicity, we set the perfect CSI case as the stage to introduce the extensions.

\subsubsection{SRRM with external eavesdroppers}
One can also consider including $L$ external eavesdroppers (Eves) into the system model. The only difference lies in the expression of the achievable secrecy rate, both the multicast message and the confidential message should be kept perfectly secure from the Eves. To put into context, let ${{\mathbf{g}}_l}\in {{\mathbb{C}}^{{1\times {N}_{t}}}}$ be the channel vector between the transmitter and Eve $l$, the achievable secrecy rate should be rewritten as
\begin{equation}
  {R_c} \le {C_b}-\mathop {\max }\{\mathop {\max }\limits_{k \in {{\cal K}_e}} {C_{e,k}},\mathop {\max }\limits_{l \in {{\cal L}_e}} {R_{e,l}},\}
\end{equation}
in which ${{\cal L}_e}=\{1,2,\cdots,L\}$ denotes the indices of the external Eves and ${R_{e,l}}=\log \left( {1 + \frac{{{{\bf{g}}_l}({{\bf{Q}}_c}+{{\bf{Q}}_0}){\bf{g}}_l^H}}{{1 + {{\bf{g}}_l}{{\bf{Q}}_a}{\bf{g}}_l^H}}} \right)$. It can be proved that the QoMS-based scalarization method is also applicable to this scenario, but with more judicious construction method to prove Property \ref{property1}. For simplicity, we omit the detailed process in this paper. The resulting scalar problem can once again be tackled using the Charnes-Cooper transformation-based line search method. Apparently, the introduction of external Eves would suppress the size of the secrecy rate region.

\subsubsection{Colluding Unauthorized Receivers}
Consider the case where the unauthorized receivers collude to collectively decode the confidential message in $J$ groups. Let ${{\mathbf{G}}_j}\in {{\mathbb{C}}^{{{N}_{c,j} \times {N}_{t}}}}$ be the channel matrix between the transmitter and the $j$th colluding group, with ${N}_{c,j}$ being the number of unauthorized receivers in $j$th colluding group. The channel matrix ${{\mathbf{G}}_j}$ is formed by stacking the channel vectors of the unauthorized receivers in $j$th colluding group. The only difference of this colluding scenario still lies in the expression of the achievable secrecy rate, i.e.,
\begin{equation}
  {R_c} \le {C_b}-\mathop {\max }\limits_{j \in {\cal J}} {R_{e,j}},
\end{equation}
in which ${\cal J}=\{1,2,\cdots,J\}$ and ${R_{e,j}}={\log \det ({\bf{I}} + {{({\bf{I}} + {{\bf{G}}_j}{{\bf{Q}}_a}{\bf{G}}_j^H)}^{ - 1}}{{\bf{G}}_j}{{\bf{Q}}_c}{\bf{G}}_j^H)}$. Though the determinant expression is generally intractable to handle, it can be tightly relaxed into a linear expression by following the approach proposed in \cite{li2013spatially}. The remnant work is to follow the same derivations as those in the case with external Eves, and the details are omitted here.

\section{Numerical Results}
In this section, we provide numerical results to illustrate the secrecy rate regions derived from our proposed optimal and suboptimal schemes, compared to some other existing schemes. The first one is the no-AN scheme, i.e., with prefixing ${{\bf{Q}}_a}$ as $\mathbf{0}$ in the primal SRRM problems. Another one is based on the traditional service integration strategies, which assign the confidential message and multicast message to two different logic channels, for instance, two orthogonal time slots. This time division multiple address (TDMA)-based service integration splits the primal SRRM problems into two conventional rate maximization problem, i.e., the SRM problem (setting $\tau _{ms}=0$) and multicast rate maximization problem (cf. (\ref{max_tau}) and (\ref{max_tau2})). For the fairness of comparison, the secrecy rate and multicast rate achieved by the TDMA scheme should be halved \cite{Wyrembelski2012Physical}. For the imperfect CSI case, we also give the secrecy rate regions achieved by a nonrobust (naive) scheme, the details of which will be introduced thereinafter. We will first consider the perfect CSI case in the first subsection, and then the imperfect CSI case in the following subsection.

\subsection{The Perfect CSI Case}
Unless specified, the simulation settings are as follows. The number of transmit antennas at the transmitter is $N_t=2$. The number of receivers is $K=5$. In the simulation, we investigate the secrecy rate regions achieved by deterministic channels, as \cite{Hung2010Multiple,liu2013new,ekrem2012capacity} did. All channels are generated from i.i.d. complex Gaussian distribution with zero mean and unit variance. In particular, the channel vectors we use are given by
\begin{equation}\label{channels}
\begin{split}
&{{\bf{h}}_1} = \left[ {\begin{array}{*{20}{c}}
{0.3802 - 1.5972i}&{1.2968 + 0.6096i}
\end{array}} \right],\\
&{{\bf{h}}_2} = \left[ {\begin{array}{*{20}{c}}
{0.2254 - 0.3066i}&{-0.9247 + 0.2423i}
\end{array}} \right],\\
&{{\bf{h}}_3} = \left[ {\begin{array}{*{20}{c}}
{0.5303 - 0.9545i}&{1.9583 + 2.1460i}
\end{array}} \right],\\
&{{\bf{h}}_4} = \left[ {\begin{array}{*{20}{c}}
{0.5129 + 0.5054i}&{-0.0446 - 0.1449i}
\end{array}} \right],\\
&{{\bf{h}}_5} = \left[ {\begin{array}{*{20}{c}}
{0.0878 - 0.9963i}&{1.0534 + 1.0021i}
\end{array}} \right],
\end{split}
\end{equation}
where $i \buildrel \Delta \over = \sqrt { - 1}$.

\begin{figure}[!t]
\centering
\includegraphics[width=2.8in]{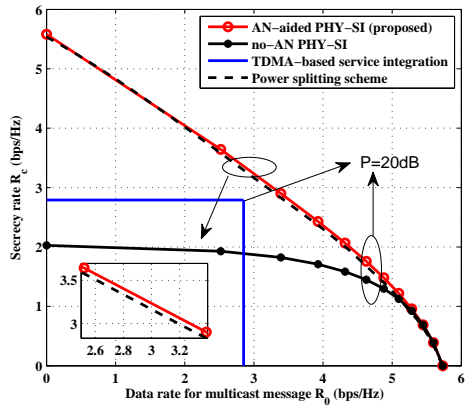}
\DeclareGraphicsExtensions.
\caption{Secrecy rate regions with perfect CSI}\label{SRR_pf}
\end{figure}
Fig.\,\ref{SRR_pf} plots the secrecy rate regions achieved by our considered schemes with $P=20$dB. The curves in Fig.\,\ref{SRR_pf} are the boundary lines of the secrecy rate regions. First, let us concentrate on the comparison between our proposed scheme and the no-AN scheme. As seen, secrecy rates with AN are mostly higher than those without AN. The striking gap indicates that AN indeed enhances the security performance without compromising the QoMS. Nonetheless, with the increasing demand for QoMS, the two curves tend to be coincident, which implies that AN is prohibitive at high QoMS region. The prohibition of AN reveals an inherent difference between PHY-SI and PHY-security: the use of AN must be more prudent due to the demand for QoMS. Next, we pay our attention to the secrecy rate region achieved by the TDMA-based scheme. As expected, our proposed scheme yields a significantly larger region than the TDMA-based one, which implies the inherent advantage of PHY-SI over traditional service integration. Finally, we can observe that the performance gap between the power splitting suboptimal scheme and the real secrecy rate region is negligible. This observation demonstrates that the power splitting scheme can achieve a near-optimal performance with higher implementation efficiency.

\begin{figure}[!t]
\centering
\includegraphics[width=2.8in]{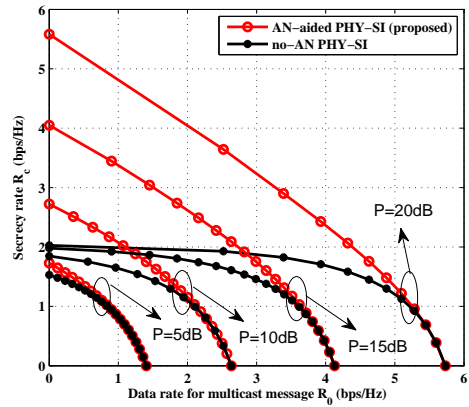}
\DeclareGraphicsExtensions.
\caption{Secrecy rate regions versus the transmit power}\label{SRR_pf_Power}
\end{figure}
Next, we pay our attention to the effect of transmit power on the achievable secrecy rate regions. Meanwhile, we plot the secrecy rate region achieved by the no-AN scheme as a benchmark. We examine four cases, namely, $P=5,10,15$ and 20dB. From Fig.\,\ref{SRR_pf_Power}, we can have some useful observations. First, our AN-aided scheme achieves a secrecy rate region larger than the no-AN one, even under low transmit power. However, the gap between these two strategies dramatically reduced when $P$ diminishes. This is due to AN's dual role in PHY-SI, i.e., in order to guarantee the QoMS, AN must decrease to reduce the interference at all receivers. The second observation is that the secrecy rate regions with AN expand more strikingly when $P$ increases. On the contrary, the secrecy rate regions without AN practically expand in the horizontal direction. That is, for the no-AN scheme, the increasing transmit power mainly contributes to the multicast message transmission, rather than the confidential message transmission. This phenomenon can be interpreted from the transmit degree of freedom (d.o.f.). The total d.o.f. of unauthorized receivers is $K-1=4$, higher than the transmit d.o.f. $N_t=2$. The lack of transmit d.o.f. is the reason for the unsatisfactory security performance of the no-AN SRRM design.

\subsection{The Imperfect CSI Case}
The simulation settings in the imperfect CSI case are generally the same as those in the perfect CSI case. The estimated channel vectors $\{{{\bf{\tilde h}}_k}\}_{k \in \cal {K}}$ are set identical to the deterministic complex channel vectors adopted in the last subsection. Without loss of generality, we set ${\varepsilon _k}=\varepsilon=0.2$ for all $k$. In the imperfect CSI case, we consider a nonrobust transmit design, and plot its achieved secrecy rate regions. Its idea is to apply the presumed CSI, $\{{{\bf{\tilde h}}_k}\}_{k \in \cal {K}}$, to perform the transmit design (cf. SRRM problem (\ref{op4})).

\begin{figure}[!t]
\centering
\includegraphics[width=2.8in]{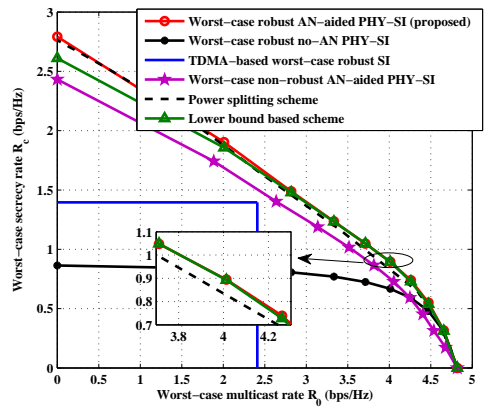}
\DeclareGraphicsExtensions.
\caption{Worst-case robust secrecy rate regions}\label{SRR_ipf_WC}
\end{figure}
We still first evaluate the resultant worst-case secrecy rate regions achieved by different schemes in Fig.\,\ref{SRR_ipf_WC}. We can clearly observe that the existence of channel uncertainty dramatically diminishes the achievable secrecy rate regions by comparing Fig.\,\ref{SRR_ipf_WC} with Fig.\,\ref{SRR_pf}. The basic observations from Fig.\,\ref{SRR_ipf_WC} is virtually similar to those from Fig.\,\ref{SRR_pf}, for example, the best performance of our proposed AN-aided scheme and the coincidence of the AN-aided scheme and the no-AN scheme at high QoMS region. Particularly, our proposed AN-aided scheme outperforms the nonrobust scheme, though the nonrobust scheme achieves a larger secrecy rate region than the no-AN one. This confirms that incorporating AN is a powerful means to combat channel uncertainties, even with integrated services. Also, we should mention that our proposed two suboptimal schemes achieve good approximation accuracies to the optimal secrecy rate region. Especially, the lower bound based scheme even yields higher secrecy rates at high QoMS region than the power splitting scheme.

\begin{figure}[!t]
\centering
\includegraphics[width=2.8in]{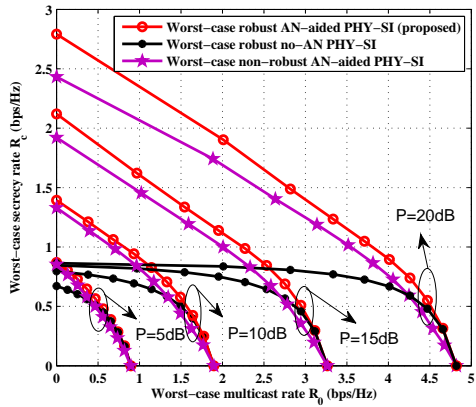}
\DeclareGraphicsExtensions.
\caption{Worst-case robust secrecy rate regions versus transmit power}\label{SRR_ipf_Power}
\end{figure}
Fig.\,\ref{SRR_ipf_Power} plots the worst-case secrecy rate regions against the transmit power. As seen, the gaps between the AN-aided and no-AN schemes have been more remarkable than those in the perfect CSI case. Besides, the d.o.f. bottleneck suffered by the no-AN design still exists in the imperfect CSI case, and becomes even more severe. Specifically, in the low QoMS region, the no-AN scheme can only attain a maximum secrecy rate of 0.8 bps/Hz with $P=20$dB. As a reminder, the robust scheme outperforms the nonrobust one over the whole range of powers tested.

\begin{figure}[!t]
\centering
\includegraphics[width=2.8in]{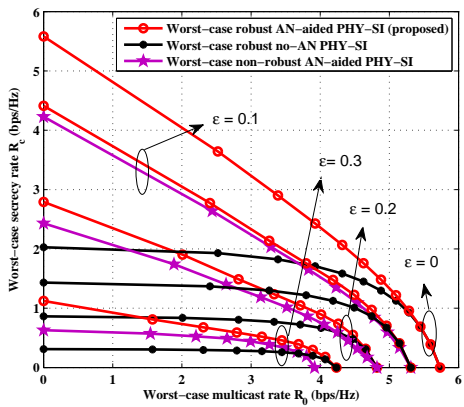}
\DeclareGraphicsExtensions.
\caption{Worst-case robust secrecy rate regions versus CSI uncertainty}\label{SRR_ipf_evar}
\vspace*{-3pt}
\end{figure}
Finally, we investigate the relation between the worst-case secrecy rate regions and the CSI uncertainty level by fixing $P=20$dB. Our benchmark is the nonrobust scheme. The results are shown in Fig.\,\ref{SRR_ipf_evar}. As expected, the basic trend is that the larger CSI uncertainties are, the smaller the worst-case secrecy rate regions are. Besides, when the channel uncertainty level $\varepsilon$ increases, the robustness of the AN-aided scheme becomes more obvious. When $\varepsilon =0.2$, the nonrobust scheme achieves a maximum multicast rate comparable to the AN-aided one. However, when $\varepsilon =0.3$, its maximum achievable multicast rate becomes smaller than the AN-aided one, and the performance gap between these two schemes expands. This phenomenon reveals the sensitivity of the nonrobust scheme to channel uncertainties, since its design can only guarantee the optimality to the presumed CSI, but not to the actual CSI.

\section{Conclusion}
In this paper, we considered an AN-aided transmit design for multiuser MISO broadcast channel with amalgamating confidential service and multicast service, with both perfect and imperfect CSI. The input covariances for confidential message, multicast message and AN were designed to maximize the achievable secrecy rate region, which is a vector maximization problem. Since the vector optimization problem is inherently complex to solve, we proved that this SRRM problem is equivalent to a standard scalar maximization problem, essentially an SRM problem with QoMS constraints. Even so, this scalar maximization problem was still hard to solve due to its non-convexity. We therefore developed an SDP-based approach to solve the problem by first introducing a two-stage reexpression. Then we showed that, for the perfect CSI case and its worst-case robust counterpart, the equivalent SRM problem can be efficiently tackled by solving a sequence of SDPs. Moreover, we proved the optimality of transmit beamforming to the confidential message transmission, and gave the complexity analysis of our proposed optimization methods. To mitigate the computational complexity, two suboptimal schemes were also proposed.

Numerical results demonstrated that our proposed AN-aided scheme always achieves larger secrecy rate regions than some other existing schemes. These observations verified the efficacy of AN in expanding the secrecy rate region, as well as the inherent advantage of PHY-SI over traditional service integration. Moreover, the results also indicated that our proposed suboptimal schemes could achieve near-optimal performance, with significant time saving. As a future direction, it would be interesting to study the combination of confidential broadcasting and multicast services.

\appendix
\subsection{Proof of Proposition \ref{proposition1}}\label{rank_proof1}
The proof is composed of two steps. First, given a feasible $\alpha$ of (\ref{op4.1}), defining the optimal objective value of (\ref{op4.2}) as ${{\bar \eta }_\alpha }$, we show that (\ref{op4.2}) has identical optimal solutions to a power minimization problem given by
\begin{subequations}\label{opA1}
\begin{align}
\nonumber&\mathop {\min }\limits_{{{\bf{Q}}_0},{{\bf{Q}}_a},{{\bf{Q}}_c}}\text{Tr}({{\bf{Q}}_0} + {{\bf{Q}}_a} + {{\bf{Q}}_c})\\
\text{s.t.}\;&\frac{{1 + {{\bf{h}}_1}({{\bf{Q}}_c} + {{\bf{Q}}_a}){\bf{h}}_1^H}}{{\alpha (1 + {{\bf{h}}_1}{{\bf{Q}}_a}{\bf{h}}_1^H)}} \ge {{\bar \eta }_\alpha },\label{opA1a}\\
&\text{(\ref{op4.2a}), (\ref{op4.2b}) and (\ref{op4.2d}) satisfied}.
\end{align}
\end{subequations}
Second, we show ${\rm{rank}}({\bf{Q}}_c^{*}) = 1$ by studying the Karush-Kuhn-Tucker (KKT) conditions of (\ref{opA1}).

\textbf{Step 1}: Assume that the optimal solutions of (\ref{op4.2}) and (\ref{opA1}) are denoted as $({{\bf{\bar Q}}_0},{{\bf{\bar Q}}_c},{{\bf{\bar Q}}_a})$ and $({{\bf{\tilde Q}}_0},{{\bf{\tilde Q}}_c},{{\bf{\tilde Q}}_a})$, respectively. One can easily verify that $({{\bf{\bar Q}}_0},{{\bf{\bar Q}}_c},{{\bf{\bar Q}}_a})$ is a feasible solution of (\ref{opA1}), which yields
\begin{equation}\label{eq3}
\text{Tr}({{\bf{\tilde Q}}_0} + {{\bf{\tilde Q}}_a} + {{\bf{\tilde Q}}_c}) \le \text{Tr}({{\bf{\bar Q}}_0} + {{\bf{\bar Q}}_a} + {{\bf{\bar Q}}_c}) \le P.
\end{equation}
The first inequality is due to the fact that any feasible solution of (\ref{opA1}) is doomed to consume no less power than that consumed by the optimal solution of (\ref{opA1}); the second inequality is owing to the fact that $({{\bf{\bar Q}}_0},{{\bf{\bar Q}}_c},{{\bf{\bar Q}}_a})$ should follow the sum power constraint in the inner maximization problem of (\ref{op4.2}).

The inequality in (\ref{eq3}) implies that $({{\bf{\tilde Q}}_0},{{\bf{\tilde Q}}_c},{{\bf{\tilde Q}}_a})$ is a feasible solution of (\ref{op4.2}). Hence, we have
\begin{equation}\label{eq4}
\frac{{1 + {{\bf{h}}_1}({{{\bf{\tilde Q}}}_c} + {{{\bf{\tilde Q}}}_a}){\bf{h}}_1^H}}{{\alpha (1 + {{\bf{h}}_1}{{{\bf{\tilde Q}}}_a}{\bf{h}}_1^H)}} \le {\bar \eta _\alpha }.
\end{equation}
Combining (\ref{opA1a}) with (\ref{eq4}), we obtain
\begin{equation}\label{eq5}
\frac{{1 + {{\bf{h}}_1}({{{\bf{\tilde Q}}}_c} + {{{\bf{\tilde Q}}}_a}){\bf{h}}_1^H}}{{\alpha (1 + {{\bf{h}}_1}{{{\bf{\tilde Q}}}_a}{\bf{h}}_1^H)}} = {\bar \eta _\alpha },
\end{equation}
which proves $({{\bf{\tilde Q}}_0},{{\bf{\tilde Q}}_c},{{\bf{\tilde Q}}_a})$ is also an optimal solution of (\ref{op4.2}).

\textbf{Step 2}: Rewrite (\ref{opA1a}) as ${{\bf{h}}_1}({{\bf{Q}}_c} + \mu {{\bf{Q}}_a}){\bf{h}}_1^H + \mu  \ge 0$, where $\mu  \buildrel \Delta \over = 1 - \alpha {\bar \eta _\alpha }$. The Lagrangian of (\ref{opA1}) is
\begin{equation}\label{Lag}
\begin{split}
&L({{\bf{Q}}_0},{{\bf{Q}}_a},{{\bf{Q}}_c},\lambda ,{\mathbf{\eta }},{\mathbf{\sigma }},{\bf{A}},{\bf{B}},{\bf{C}}) = \\ &\text{Tr}({{\bf{Q}}_0} + {{\bf{Q}}_a} + {{\bf{Q}}_c})- \lambda [{{\bf{h}}_1}({{\bf{Q}}_c} + \mu {{\bf{Q}}_a}){\bf{h}}_1^H + \mu ]\\
&- \sum\limits_{k = 2}^K {{\eta _k}[(\alpha  - 1)(1 + {{\bf{h}}_k}{{\bf{Q}}_a}{\bf{h}}_k^H) - {{\bf{h}}_k}{{\bf{Q}}_c}{\bf{h}}_k^H]}   \\
&-\sum\limits_{k = 1}^K {{\sigma _k}[{{\bf{h}}_k}{{\bf{Q}}_0}{\bf{h}}_k^H - \tau '{{\bf{h}}_k}({{\bf{Q}}_a} + {{\bf{Q}}_c}){\bf{h}}_k^H  - \tau ']}  \\
& -\text{Tr}({\bf{A}}{{\bf{Q}}_a})-\text{Tr}({\bf{B}}{{\bf{Q}}_0}) -\text{Tr}({\bf{C}}{{\bf{Q}}_c}),
\end{split}
\end{equation}
where ${\bf{A}} \succeq {\bf{0}}, {\bf{B}} \succeq {\bf{0}}, {\bf{C}} \succeq {\bf{0}}, \lambda > 0,{\eta _k} \ge 0,\forall k \in {{\cal K}_e}$ and ${\sigma _k} \ge 0,\forall k \in {\cal K}$ are dual variables pertaining to primal constraints in (\ref{opA1}). To prove ${\rm{rank}}({{\bf{\tilde Q}}_c}) = 1$, we pick up the following KKT conditions to check.
\begin{subequations}\label{kkt}
\begin{align}
{\bf{T}} - \lambda {\bf{h}}_1^H{{\bf{h}}_1} = {\bf{C}},\label{kkt1}\\
{\bf{C}}{{\bf{\tilde Q}}_c} = {\bf{0}},\label{kkt2}\\
{\eta _k} \ge 0,\forall k \in {{\cal K}_e},\label{kkt4}\\
{\sigma _k} \ge 0,\forall k \in {\cal K},\label{kkt5}
\end{align}
\end{subequations}
in which ${\bf{T}} \buildrel \Delta \over = {\bf{I}} + \sum\limits_{k = 2}^K {{\eta _k}{\bf{h}}_k^H{{\bf{h}}_k}}  + \tau '\sum\limits_{k = 1}^K {{\sigma _k}{\bf{h}}_k^H{{\bf{h}}_k}}$. Combining (\ref{kkt1}) with (\ref{kkt2}) yields ${\bf{T}}{{\bf{\tilde Q}}_c} = \lambda {\bf{h}}_1^H{{\bf{h}}_1}{{\bf{\tilde Q}}_c}$, and we know ${\bf{T}} \succ {\bf{0}}$ from (\ref{kkt4}) and (\ref{kkt5}), one can obtain
\begin{equation}\label{eq7}
{\rm{rank}}({{\bf{\tilde Q}}_c}) = {\rm{rank}}(\lambda {\bf{h}}_1^H{{\bf{h}}_1}{{\bf{\tilde Q}}_c}) \le 1,
\end{equation}
which implies that ${\rm{rank}}({{\bf{\tilde Q}}_c}) \le 1$ holds for any feasible $\alpha$ of (\ref{op4.1}). Eliminating the trivial solution ${\bf{\tilde Q}}_c = {\bf{0}}$,  we obtain ${\rm{rank}}({\bf{\tilde Q}}_c) = 1$.

\subsection{Proof of Proposition \ref{CA}}\label{CA_proof1}
Suppose that $\alpha^*$ is an optimal solution of problem (\ref{op4.1}), and that $\left( {{{\bf{Z}}^*},{{\bf{\Gamma }}^*},{{\bf{\Phi }}^*},{\xi^*}} \right)$ is an optimal solution of problem (\ref{op5}). For any $\Delta >0$ such that ${\alpha ^ * } + \Delta \in [1,1 + P{\left\| {{{\bf{h}}_1}} \right\|^2}]$, we must have
\begin{equation}\label{ca1}
\log \left(\eta (\tau ',\alpha^* )\right) \ge \log \left(\eta (\tau ',\alpha^* + \Delta)\right).
\end{equation}
For ease of exposition, the dependence of $\eta$ on $\tau '$ will be omitted in the following proof of Proposition \ref{CA}.

Consider the function $\eta (\alpha^* + \Delta)$, that is,
\begin{subequations}\label{ca2}
\begin{align}
&\nonumber\eta (\alpha^* + \Delta) = \mathop {\max}\limits_{{\bf{Z}},{\bf{\Gamma }},{\bf{\Phi }},\xi }\xi  + {{\bf{h}}_1}({\bf{Z}} + {\bf{\Gamma }}){\bf{h}}_1^H\\
\text{s.t.}\quad &\xi  + {{\bf{h}}_1}{\bf{\Gamma h}}_1^H = (\alpha^* + \Delta)^{-1},\\
&(\alpha^* + \Delta  - 1)(\xi  + {{\bf{h}}_k}{\bf{\Gamma h}}_k^H) \ge {{\bf{h}}_k}{\bf{Zh}}_k^H,\forall k \in {{\cal K}_e}.\\
&\text{(\ref{op5c})-(\ref{op5e}) satisfied}.
\end{align}
\end{subequations}
Let $p = \frac{\alpha^* }{{\alpha^*  + \Delta }}$, and $( {{\bf{\hat Z}},{\bf{\hat \Gamma }},{\bf{\hat \Phi }},{\hat \xi } } ) = p\left( {{\bf{Z}}^*,{\bf{\Gamma }}^*,{\bf{\Phi }}^*,\xi^* } \right)$. One can easily check that $( {{\bf{\hat Z}},{\bf{\hat \Gamma }},{\bf{\hat \Phi }},\hat \xi } )$ is feasible to (\ref{ca2}). Accordingly, we obtain
\begin{equation}\label{ca3}
\begin{split}
p\eta (\alpha^* ) &= p({\xi^*}  + {{\bf{h}}_1}({{\bf{Z}}^*} + {\bf{\Gamma }^*}){\bf{h}}_1^H)\\
&={\hat \xi } + {{\bf{h}}_1}({\bf{\hat Z}} + {\bf{\hat \Gamma }}){\bf{h}}_1^H \\
&\le \eta (\alpha^* + \Delta),
\end{split}
\end{equation}
in which the first inequality is due to the optimality of $\left( {{{\bf{Z}}^*},{{\bf{\Gamma }}^*},{{\bf{\Phi }}^*},{\xi^*}} \right)$ to (\ref{op5}), while the last inequality is resulted from the feasibility of $({{\bf{\hat Z}},{\bf{\hat \Gamma }},{\bf{\hat \Phi }},\hat \xi })$ to (\ref{ca2}).

Our next step is to characterize the rate gap between $\log \left(\eta (\alpha^* )\right)$ and $\log \left(\eta (\alpha^* + \Delta)\right)$, i.e.,
\begin{equation}\label{ca4}
\begin{split}
\log \left(\eta (\alpha^* )\right) - \log \left(\eta (\alpha^* + \Delta)\right)&= \log \left(\frac{{\eta (\alpha^* )}}{{\eta (\alpha^*  + \Delta )}}\right),\\
&\le \log \left(\frac{1}{p}\right),
\end{split}
\end{equation}
in which the last inequality is derived from (\ref{ca3}). In order to obtain an $\epsilon$-suboptimal solution $\alpha^*  + \Delta$, we set
\begin{equation}\label{ca5}
\log \left(\frac{1}{p}\right)< \epsilon,
\end{equation}
which can be simplified as $\Delta  < {\alpha ^*}({2^\epsilon } - 1)$, and we choose
\begin{equation}\label{ca6}
\Delta  = {2^\epsilon } - 1.
\end{equation}
Therefore, when uniform sampling search is adopted, the maximum number of searches for one boundary point is
\begin{equation}\label{ca7}
T_1 = \frac{{(1 + P{{\left\| {{{\bf{h}}_1}} \right\|}^2}) - 1}}{\Delta } = \frac{{P{{\left\| {{{\bf{h}}_1}} \right\|}^2}}}{{{2^\epsilon } - 1}}.
\end{equation}

Regarding the inner SDP problem (\ref{op5}), it involves 3 LMI constraints of size $N_t$, and $2K+1$ LMI constraints of size 1. As a consequence, when a standard interior-point method (IPM) is used, the resultant arithmetic computation cost of solving (\ref{op5}) should be on the order of $\ln{(1/\epsilon)}\sqrt \gamma  \zeta$ \cite[Lecture 6]{ben2001lectures}, where $\gamma$ and $\zeta$ is given in (\ref{comp.order}). This fact completes the proof.

\subsection{Proof of Proposition \ref{proposition4}}\label{rank_proof2}
The proof is composed of two steps. First, given a feasible $\beta$ of (\ref{op9.1}), defining the optimal objective value of (\ref{op9.2}) as ${{\bar \eta }_\beta }$, we consider the following power minimization problem, i.e.,
\begin{subequations}\label{opA2}
\begin{align}
\nonumber&\mathop {\min }\limits_{{{\bf{Q}}_0},{{\bf{Q}}_a},{{\bf{Q}}_c},\atop
{\left\{ {{t_k}} \right\}_{k \in {{\cal K}_e}}},{\left\{ {{\delta _k}} \right\}_{k \in {\cal K}}}} {\rm{Tr}}({{\bf{Q}}_0} + {{\bf{Q}}_a} + {{\bf{Q}}_c})\\
\text{s.t.}\quad &\mathop {\min }\limits_{{{\bf{h}}_1} \in {B_1}} \frac{{1 + {{\bf{h}}_1}({{\bf{Q}}_c} + {{\bf{Q}}_a}){\bf{h}}_1^H}}{{\beta (1 + {{\bf{h}}_1}{{\bf{Q}}_a}{\bf{h}}_1^H)}} \ge {{\bar \eta }_\beta },\label{opA2a}\\
&{{\bf{T}}_k}(\beta ,{{\bf{Q}}_c},{{\bf{Q}}_a},{t_k}) \succeq {\bf{0}},{t_k} \ge 0,\forall k \in {{\cal K}_e},\\
&{{\bf{S}}_k}(\tau ',{{\bf{Q}}_c},{{\bf{Q}}_a},{{\bf{Q}}_0},{\delta _k}) \succeq {\bf{0}},{\delta _k} \ge 0,\forall k \in {\cal K},\\
&{{\bf{Q}}_0} \succeq {\bf{0}}, {{\bf{Q}}_a} \succeq {\bf{0}}, {{\bf{Q}}_c} \succeq {\bf{0}}.
\end{align}
\end{subequations}
Following the same procedures in the proof of Proposition \ref{proposition1}, it is easy to verify that the optimal solution of (\ref{opA2}), denoted by $({{\bf{\tilde Q}}_0},{{\bf{\tilde Q}}_c},{{\bf{\tilde Q}}_a})$, must be optimal for (\ref{op9.2}). Second, we will prove that ${\rm{rank}}({{\bf{\tilde Q}}_c}) \le 1$ by checking the KKT conditions of (\ref{opA2}).

Define ${{\bf{\hat h}}_k} = {[{\bf{I}},{\bf{\tilde h}}_k^H]^H}, \forall k \in {\cal K}$. By using ${\cal S}$-procedure, we first reformulate (\ref{opA2a}) as
\begin{equation}\label{eq8}
{\bf{U}}(\beta ,{{\bf{Q}}_c},{{\bf{Q}}_a},\rho ) \buildrel \Delta \over = {{{\bf{\hat h}}}_1}({{\bf{Q}}_c} + (1 - \beta {{\bar \eta }_\beta }){{\bf{Q}}_a}){\bf{\hat h}}_1^H + \bf \Xi,
\end{equation}
in which $\bf \Xi = \left[ {\begin{array}{*{20}{c}}
{\rho {\bf{I}}}&{\bf{0}}\\
{\bf{0}}&{1 - \rho \varepsilon _1^2 - {\beta {{\bar \eta }_\beta }}}
\end{array}} \right]$ and $\rho \ge 0$, and then rewrite ${{\bf{T}}_k}$ and ${{\bf{S}}_k}$ in (\ref{opA2}) as the following form.
\begin{align}
&{{\bf{T}}_k}={{{\bf{\hat h}}}_k}((\beta  - 1){{\bf{Q}}_a} - {{\bf{Q}}_c}){\bf{\hat h}}_k^H + \bf \Upsilon, \nonumber\\
&{{\bf{S}}_k}={{{\bf{\hat h}}}_k}({{\bf{Q}}_0} - \tau '({{\bf{Q}}_a} + {{\bf{Q}}_c})){\bf{\hat h}}_k^H + \bf\Omega,\label{opA3}
\end{align}
where \[\bf\Upsilon = \left[ {\begin{array}{*{20}{c}}
{{t_k}{\bf{I}}}&{\bf{0}}\\
{\bf{0}}&{ - {t_k}\varepsilon _k^2 + \beta  - 1}
\end{array}} \right], \bf\Omega =\left[ {\begin{array}{*{20}{c}}
{{\delta _k}{\bf{I}}}&{\bf{0}}\\
{\bf{0}}&{ - {\delta _k}\varepsilon _k^2 - \tau '}
\end{array}} \right].\] The Lagrangian associated with (\ref{opA2}) is therefore given by
\begin{figure*}[!b]
\normalsize
\setcounter{MYtempeqncnt}{\value{equation}}
\setcounter{equation}{72}
\vspace*{-5pt}
\hrulefill
\vspace*{12pt}
\begin{subequations}\label{wcca3}
\begin{align}
\nonumber\eta(\tau',\beta)=&\mathop {\max }\limits_{{\mathbf{Z}},{\bf{\Gamma}},{\mathbf{\Phi }},\xi\atop {\left\{ {{\lambda_k}} \right\}_{k \in {{\cal K}_e}}},{\left\{ {{\mu _k}} \right\}_{k \in {\cal K}}}} \mathop {\min }\limits_{{{\bf{h}}_1} \in {B_1}} \frac{{\xi + {{\bf{h}}_1}({\mathbf{Z}} + {\mathbf{\Gamma }}){\bf{h}}_1^H}}{{\beta(\xi + {{\bf{h}}_1}{\mathbf{\Gamma }}{\bf{h}}_1^H)}}\\
\text{s.t.}\quad &{{\bf{\tilde T}}_k}(\beta ,{\mathbf{Z}},{\mathbf{\Gamma }},{\lambda_k}) = \left[ {\begin{array}{*{20}{c}}
{{\lambda_k}{\bf{I}} + (\beta  - 1){\mathbf{\Gamma }} - {\mathbf{Z}}}&{((\beta  - 1){\mathbf{\Gamma }} - {\mathbf{Z}}){\bf{\tilde h}}_k^H}\\
{{{{\bf{\tilde h}}}_k}((\beta  - 1){\mathbf{\Gamma }} - {\mathbf{Z}})}&{{{{\bf{\tilde h}}}_k}((\beta  - 1){\mathbf{\Gamma }} - {\mathbf{Z}}){\bf{\tilde h}}_k^H - {\lambda_k}\varepsilon _k^2 + (\beta  - 1)\xi}
\end{array}} \right] \succeq {\bf{0}}, \\
&{{\bf{\tilde S}}_k}({\mathbf{Z}},{\mathbf{\Gamma }},{\mathbf{\Phi }},{\mu_k}) = \left[ {\begin{array}{*{20}{c}}
{{\mu _k}{\bf{I}} + {\mathbf{\Phi }} - \tau '({\mathbf{\Gamma }} + {\mathbf{Z}})}&{({\mathbf{\Phi }} - \tau '({\mathbf{\Gamma }} + {\mathbf{Z}})){\bf{\tilde h}}_k^H}\\
{{{{\bf{\tilde h}}}_k}({\mathbf{\Phi }} - \tau '({\mathbf{\Gamma }} + {\mathbf{Z}}))}&{- {\mu _k}\varepsilon _k^2 - \tau '\xi + {{{\bf{\tilde h}}}_k}({\mathbf{\Phi }} - \tau '({\mathbf{\Gamma }} + {\mathbf{Z}})){\bf{\tilde h}}_k^H}
\end{array}} \right] \succeq {\bf{0}},\forall k \in {\cal K}, \label{wcca3b}\\
&\text{Tr}({\bf{\Phi }} + {\bf{\Gamma }} + {\bf{Z}}) \le P\xi ,\label{wcca3c}\\
&{{\bf{Z}}} \succeq {\bf{0}}, {\bf{\Gamma }} \succeq {\bf{0}}, {\bf{\Phi }} \succeq {\bf{0}}, {\lambda_k} \ge 0,\forall k \in {{\cal K}_e}, {\mu_k} \ge 0,\forall k \in {\cal K}.\label{wcca3d}
\end{align}
\end{subequations}
\setcounter{equation}{\value{MYtempeqncnt}}
\vspace*{-12pt}
\end{figure*}
\begin{equation}\label{Lag}
\begin{split}
&L({\bf{X}}) = {\rm{Tr}}({{\bf{Q}}_0} + {{\bf{Q}}_a} + {{\bf{Q}}_c}) - {\rm{Tr}}({\bf{\Phi U}}(\beta ,{{\bf{Q}}_c},{{\bf{Q}}_a},\rho ))\\
&- \sum\limits_{k \in {{\cal K}_e}} {{\rm{Tr}}({{\bf{\Psi }}_k}{{\bf{T}}_k}(\beta ,{{\bf{Q}}_c},{{\bf{Q}}_a},{t_k}))} \\
&- \sum\limits_{k \in {\cal K}} {{\rm{Tr}}({{\bf{\Lambda }}_k}{{\bf{S}}_k}(\tau ',{{\bf{Q}}_c},{{\bf{Q}}_a},{{\bf{Q}}_0},{\delta _k}))} \\
&- {\rm{Tr}}({\bf{A}}{{\bf{Q}}_a}) - {\rm{Tr}}({\bf{B}}{{\bf{Q}}_0}) - {\rm{Tr}}({\bf{C}}{{\bf{Q}}_c})\\
&- \sum\limits_{k \in {{\cal K}_e}} {{\eta _k}{t_k}}  - \sum\limits_{k \in {\cal K}} {{\upsilon _k}{\delta _k}} - \sigma \rho ,
\end{split}
\end{equation}
where $\bf{X}$ denotes a collection of all primal and dual variables: ${\bf{A}} \succ {\bf{0}},{\bf{B}} \succ {\bf{0}},{\bf{C}} \succ {\bf{0}},{\bf{\Phi }} \succ {\bf{0}},\sigma \ge 0,{{\bf{\Psi }}_k} \succ {\bf{0}},{\eta _k} \ge 0,\forall k \in {{\cal K}_e}$ and ${{\bf{\Lambda }}_k} \succ {\bf{0}},{\upsilon _k} \ge 0,\forall k \in {\cal K}$ are dual variables pertaining to primal constraints in (\ref{opA3}). To prove ${\rm{rank}}({{\bf{\tilde Q}}_c}) = 1$, we pick up the following KKT conditions to check, where we define ${\bf{R}} = {\bf{I}} + \sum\limits_{k \in {{\cal K}_e}} {{\bf{\hat h}}_k^H{{\bf{\Psi }}_k}{{{\bf{\hat h}}}_k}} + \tau '\sum\limits_{k \in {\cal K}} {{\bf{\hat h}}_k^H{{\bf{\Lambda }}_k}{{{\bf{\hat h}}}_k}}$.
\begin{subequations}\label{kkt}
\begin{align}
&{\bf{R}} - {\bf{\hat h}}_1^H{\bf{\Phi }}{{{\bf{\hat h}}}_1} = {\bf{C}}, \label{kkt6} \\
&{\bf{C}}{{\bf{\tilde Q}}_c} = {\bf{0}}, \label{kkt7} \\
&{\bf{\Phi U}}(\beta ,{{\bf{\tilde Q}}_c},{{\bf{\tilde Q}}_a},\rho )= {\bf{0}}, \label{kkt8} \\
&{{\bf{\Psi }}_k} \succeq {\bf{0}},\forall k \in {{\cal K}_e}, \label{kkt10}\\
&{{\bf{\Lambda }}_k},{\bf{\Phi }} \succeq {\bf{0}},\forall k \in {\cal K}.\label{kkt11}
\end{align}
\end{subequations}
Combining (\ref{kkt6}) with (\ref{kkt7}) yields
\begin{equation}\label{eq9}
{\bf{R}}{{\bf{\tilde Q}}_c} =  {\bf{\hat h}}_1^H{\bf{\Phi }}{{\bf{\hat h}}_1}{{\bf{\tilde Q}}_c},
\end{equation}
and we know ${\bf{R}} \succ {\bf{0}}$ from (\ref{kkt10}) and (\ref{kkt11}), one can obtain
\begin{equation}\label{eq10}
{\rm{rank}}({{\bf{\tilde Q}}_c}) = {\rm{rank}}({\bf{\hat h}}_1^H{\bf{\Phi }}{{\bf{\hat h}}_1}{{\bf{\tilde Q}}_c}) \le {\rm{rank}}({\bf{\hat h}}_1^H{\bf{\Phi }}{{\bf{\hat h}}_1}).
\end{equation}
If we can prove ${\rm{rank}}({\bf{\hat h}}_1^H{\bf{\Phi }}{{\bf{\hat h}}_1})=1$, then we will obtain ${\rm{rank}}({{\bf{\tilde Q}}_c}) \le 1$ from (\ref{eq10}). Therefore, in the remaining part of the proof, we will focus on the rank of ${\bf{\hat h}}_1^H{\bf{\Phi }}{{\bf{\hat h}}_1}$.

Substituting (\ref{eq8}) into the KKT condition (\ref{kkt8}), we obtain
\begin{equation}\label{eq11}
{\bf{\Phi}}{{{\bf{\hat h}}}_1}{\bf{\tilde Q}}{\bf{\hat h}}_1^H + {\bf{\Phi}}\bf\Xi={\bf{0}},
\end{equation}
where ${\bf{\tilde Q}} \buildrel \Delta \over = {{\bf{\tilde Q}}_c} + (1 - \beta {{\bar \eta }_\beta }){{\bf{\tilde Q}}_a}$. Premultiplying (\ref{eq11}) by ${{\bf{\hat h}}_k^H}$, we obtain
\begin{equation}\label{eq12}
{{\bf{\hat h}}_k^H}{\bf{\Phi}}{{{\bf{\hat h}}}_1}{\bf{\tilde Q}}{\bf{\hat h}}_1^H + {{\bf{\hat h}}_k^H}{\bf{\Phi}}\bf\Xi={\bf{0}}.
\end{equation}
One can easily check that
\begin{gather}\label{eq13}
\nonumber{\bf\Xi}\left[ {\begin{array}{*{20}{c}}
{{{\bf{I}}_{{N_t}}}}\\
0\end{array}} \right]=\rho \left( {{{\bf{\hat h}}}_1} - \left[ {\begin{array}{*{20}{c}}
\mathbf{0}\\{{{{\bf{\tilde h}}}_1}}
\end{array}} \right] \right)\\
{\bf{\hat h}}_1^H\left[ {\begin{array}{*{20}{c}}
{{{\bf{I}}_{{N_t}}}}\\
0\end{array}} \right] = \left[ {\begin{array}{*{20}{c}}
{{{\bf{I}}_{{N_t}}}}&{{\bf{\tilde h}}_1^H}
\end{array}} \right]\left[ {\begin{array}{*{20}{c}}
{{{\bf{I}}_{{N_t}}}}\\
0\end{array}} \right] = {{\bf{I}}_{{N_t}}},
\end{gather}
and we then postmultiply the both sides of (\ref{eq12}) by the matrix ${\left[ {\begin{array}{*{20}{c}}{{{\bf{I}}_{{N_t}}}}&0\end{array}} \right]^H}$ to get
\begin{equation}\label{eq14}
{\bf{\hat h}}_1^H{\bf{\Phi }}{{{\bf{\hat h}}}_1}{\bf{\tilde Q}} + {\bf{\hat h}}_1^H{\bf{\Phi }}\rho \left( {{{{\bf{\hat h}}}_1} - \left[ {\begin{array}{*{20}{c}}
\mathbf{0}\\{{{{\bf{\tilde h}}}_1}}
\end{array}} \right]} \right) = {\bf{0}},
\end{equation}
or equivalently,
\begin{equation}\label{eq15}
{\bf{\hat h}}_1^H{\bf{\Phi }}{{{\bf{\hat h}}}_1}({\bf{\tilde Q}} + \rho {{\bf{I}}_{{N_t}}}) = \rho {\bf{\hat h}}_1^H{\bf{\Phi }} {\left[ {\begin{array}{*{20}{c}}
\mathbf{0}\\{{{{\bf{\tilde h}}}_1}}
\end{array}} \right]}.
\end{equation}

\begin{lemma}[\cite{horn2012matrix}]\label{block}
If a block hermitian matrix ${\bf{P}} = \left[ {\begin{array}{*{20}{c}}
{{{\bf{P}}_1}}&{{{\bf{P}}_2}}\\
{{{\bf{P}}_3}}&{{{\bf{P}}_4}}
\end{array}} \right] \succeq \bf{0}$, then the main diagonal matrices ${{{\bf{P}}_1}}$ and ${{{\bf{P}}_4}}$ are always PSD matrices.
\end{lemma}

With Lemma \ref{block} and ${\bf{U}}(\beta ,{{\bf{\tilde Q}}_c},{{\bf{\tilde Q}}_a},\rho ) \succeq \bf{0}$, we can claim ${\bf{\tilde Q}} + \rho {{\bf{I}}_{{N_t}}}$ is a PSD matrix and nonsingular. Since multiplying (left/right) by a nonsingular matrix (of appropriate dimension) does not change the matrix rank, the following rank relation holds, i.e.,
\begin{equation}\label{eq16}
\begin{split}
{\rm{rank}}\left({\bf{\hat h}}_1^H{\bf{\Phi }}{{\bf{\hat h}}_1}\right)&={\rm{rank}}\left({\bf{\hat h}}_1^H{\bf{\Phi }}{{{\bf{\hat h}}}_1}({{\bf{\tilde Q}}} + \rho {{\bf{I}}_{{N_t}}})\right)\\
&={\rm{rank}}\left(\rho {\bf{\hat h}}_1^H{\bf{\Phi }} {\left[ {\begin{array}{*{20}{c}}
\mathbf{0}\\{{{{\bf{\tilde h}}}_1}}
\end{array}} \right]}\right)\\
&\le {\rm{rank}}\left({\left[ {\begin{array}{*{20}{c}}
\mathbf{0}\\{{{{\bf{\tilde h}}}_1}}
\end{array}} \right]}\right)\le 1.
\end{split}
\end{equation}

With (\ref{eq10}) and (\ref{eq16}), it is immediate to get
\begin{equation}
{\rm{rank}}({{\bf{\tilde Q}}_c}) \le {\rm{rank}}({\bf{\hat h}}_1^H{\bf{\Phi }}{{\bf{\hat h}}_1}) \le 1.
\end{equation}
Eliminating the trivial solution ${\bf{\tilde Q}}_c = {\bf{0}}$, we obtain ${\rm{rank}}({\bf{\tilde Q}}_c) = 1$.

\subsection{Proof of Proposition \ref{WCCA}}\label{CA_proof2}
First, for the quasiconcave problem in (\ref{op9.2}), its searching lower bound and upper bound can be chosen as $1/\beta$ and $\beta_{\max}/\beta$, respectively (cf. (\ref{beta_upper})). Therefore, for a given $\beta$ and a preset convergence tolerance $\epsilon_b$, the maximum number of bisection search is determined by \cite[p146]{boyd2009convex}
\begin{equation}\label{wcca1}
{M_\beta }=\log \left( {\frac{{{\beta _{\max }} - 1}}{{\beta {\epsilon _b}}}} \right).
\end{equation}

Next, we introduce the following transformation, i.e.,
\begin{equation}\label{eq17}
\begin{split}
&\xi > 0, {{\mathbf{Q}}_c} = {\mathbf{Z}}/\xi , {{\mathbf{Q}}_a} = {\mathbf{\Gamma }}/\xi , {{\mathbf{Q}}_0} = {\mathbf{\Phi }}/\xi ,\\
&{t_k} = {\lambda _k}/\xi ,\forall k \in {{\cal K}_e}, {\delta_k} = {\mu _k}/\xi ,\forall k \in {{\cal K}}
\end{split}
\end{equation}
to convert the inner quasiconcave problem (\ref{op9.2}) into the optimization problem \addtocounter{equation}{1}(\ref{wcca3}) shown at the bottom of this page. Problem (\ref{wcca3}) is still a quasiconcave maximization problem. Our purpose of introducing the transformation (\ref{eq17}) is to make the methods used in the proof of Proposition \ref{CA} applicable to the proof of Proposition \ref{WCCA}.

Suppose that $\beta^*$ is the optimal solution of problem (\ref{op9.1}), and that $\left( {{{\bf{Z}}^*},{{\bf{\Gamma }}^*},{{\bf{\Phi }}^*},{\xi^*}},{\{ {{\lambda^*_k}}\}_{k \in {{\cal K}_e}}},{\left\{ {{\mu^*_k}} \right\}_{k \in {\cal K}}} \right)$ is the optimal solution of problem (\ref{wcca3}). For any $\Delta >0$ such that ${\beta ^ * } + \Delta \in [1,\beta _{\max }]$, we must have
\begin{equation}\label{wcca2}
\log \left(\eta (\tau ',\beta^ * )\right) \ge \log \left(\eta (\tau ',\beta^ * + \Delta)\right).
\end{equation}
Again, the dependence of $\eta$ on $\tau '$ will be omitted thereinafter for brevity.

Consider the function $\eta (\beta^ * + \Delta)$, that is,
\begin{subequations}\label{wcca4}
\begin{align}
\nonumber\eta (\beta^ * + \Delta) &= \mathop {\max }\limits_{{\mathbf{Z}},{\bf{\Gamma}},{\mathbf{\Phi }},\xi\atop {\left\{ {{\lambda_k}} \right\}_{k \in {{\cal K}_e}}},{\left\{ {{\mu _k}} \right\}_{k \in {\cal K}}}} \mathop {\min }\limits_{{{\bf{h}}_1} \in {B_1}} \frac{{\xi + {{\bf{h}}_1}({\mathbf{Z}} + {\mathbf{\Gamma }}){\bf{h}}_1^H}}{{(\beta^ * + \Delta)(\xi + {{\bf{h}}_1}{\mathbf{\Gamma }}{\bf{h}}_1^H)}}\\
\text{s.t.}\; &{{\bf{\tilde T}}_k}(\beta^ * + \Delta ,{\mathbf{Z}},{\mathbf{\Gamma }},{\lambda_k}) \succeq \mathbf{0},\forall k \in {{\cal K}_e},\\
&\text{(\ref{wcca3b})-(\ref{wcca3d}) satisfied}.
\end{align}
\end{subequations}
Let $p = \frac{{\beta^*}}{\beta^* + \Delta }$, and $({{\bf{\hat Z}}},{\bf{\hat \Gamma}},{\mathbf{\hat \Phi }},{\hat \xi},{\{ {{\hat\lambda_k}}\}_{k \in {{\cal K}_e}}},{\{ {{\hat\mu_k}}\}_{k \in {\cal K}}})= p({{\bf{ Z}}^*},{\bf{\Gamma}}^*,{\mathbf{\Phi }}^*,{\xi^*},{\left\{ {{\lambda^*_k}} \right\}_{k \in {{\cal K}_e}}},{\left\{ {{\mu^*_k}} \right\}_{k \in {\cal K}}})$. One can check that $({{\bf{\hat Z}}},{\bf{\hat \Gamma}},{\mathbf{\hat \Phi }},{\hat \xi},{\{ {{\hat\lambda_k}}\}_{k \in {{\cal K}_e}}},{\{ {{\hat\mu_k}}\}_{k \in {\cal K}}})$ is feasible to (\ref{wcca4}). Accordingly, we obtain
\begin{equation}\label{wcca5}
\begin{split}
p\eta ({\beta ^ * }) &= p\mathop {\min }\limits_{{{\bf{h}}_1} \in {B_1}} \frac{{{\xi ^ * } + {{\bf{h}}_1}({{\bf{Z}}^ * } + {{\bf{\Gamma }}^ * }){\bf{h}}_1^H}}{{{\beta ^ * }({\xi ^ * } + {{\bf{h}}_1}{{\bf{\Gamma }}^ * }{\bf{h}}_1^H)}}\\
&=p\mathop {\min }\limits_{{{\bf{h}}_1} \in {B_1}} \frac{{\hat \xi  + {{\bf{h}}_1}({\bf{\hat Z}} + {\bf{\hat \Gamma }}){\bf{h}}_1^H}}{{{\beta ^ * }(\hat \xi  + {{\bf{h}}_1}{\bf{\hat \Gamma h}}_1^H)}}\\
&= \mathop {\min }\limits_{{{\bf{h}}_1} \in {B_1}} \frac{{\hat \xi  + {{\bf{h}}_1}({\bf{\hat Z}} + {\bf{\hat \Gamma }}){\bf{h}}_1^H}}{{({\beta ^ * } + \Delta )(\hat \xi  + {{\bf{h}}_1}{\bf{\hat \Gamma h}}_1^H)}}\\
&\le \eta ({\beta ^ * } + \Delta ).
\end{split}
\end{equation}
in which the first equality is due to the optimality of $({{\bf{ Z}}^*},{\bf{\Gamma}}^*,{\mathbf{\Phi }}^*,\xi^*)$ to (\ref{wcca3}), and the last inequality is due to the feasibility of $({\bf{\hat Z}},{\bf{\hat\Gamma}},{\mathbf{\hat \Phi }},\hat \xi)$ to (\ref{wcca4}). Because of the use of the bisection method, the real output of $\eta ({\beta ^ * } + \Delta )$ should be no less than $\eta ({\beta ^ * } + \Delta )- \epsilon_b$.

Our next step is to characterize the rate gap between $\log \left(\eta (\beta^* )\right)$ and $\log \left(\eta (\beta^* + \Delta)- \epsilon_b \right)$, i.e.,
\begin{equation}\label{wcca6}
\begin{split}
0 &< \log \left(\eta (\beta^* )\right) - \log \left(\eta (\beta^* + \Delta)- \epsilon_b \right)\\
&= \log \left(\frac{{\eta (\beta^* )}}{{\eta (\beta^*  + \Delta )- \epsilon_b }}\right),\\
&\mathop  \le \limits^{(a)} \log \left(\frac{{\eta (\beta^* )}}{{\frac{{{\beta ^ * }}}{{{\beta ^ * } + \Delta }}\eta (\beta^ * )- \epsilon_b }}\right),\\
&\mathop  \le \limits^{(b)} \log \left(\frac{{{\beta ^ * } + \Delta }}{{{\beta ^ * } - ({\beta ^ * } + \Delta ){\epsilon_b}}}\right),
\end{split}
\end{equation}
in which the inequality $(a)$ is derived from (\ref{wcca4}), and the inequality $(b)$ is derived from the fact $\log \eta ({\beta ^ * }) \ge 0$. In order to obtain an $\epsilon$-suboptimal solution $\beta^*  + \Delta$, we set
\begin{equation}\label{wcca7}
\log \left(\frac{{{\beta ^ * } + \Delta }}{{{\beta ^ * } - ({\beta ^ * } + \Delta ){\epsilon_b}}}\right)< \epsilon,
\end{equation}
which can be satisfied by choosing
\begin{equation}\label{wcca8}
\Delta  = \frac{{{2^\epsilon }(1 - {\epsilon _b}) - 1}}{{1 + {2^\epsilon }{\epsilon _b}}}.
\end{equation}
If $\Delta > 0$, i.e., ${\epsilon _b} < 1 - {2^{ - \epsilon }}$ is ensured, then the maximum number of uniform sampling searches could be determined by
\begin{equation}\label{wcca9}
M_u = \frac{{\beta_{\max} - 1}}{\Delta } = \frac{(1 + {2^\epsilon }{\epsilon _b})P({\lVert {{{{\bf{\tilde h}}}_1}} \rVert-{\varepsilon _1})^2}}{{{2^\varepsilon }(1 - {\epsilon _b}) - 1}}.
\end{equation}
Combining with the searching times of the bisection method, we arrive at the maximum total number of searches for one boundary point, i.e.,
\begin{equation}\label{wcca10}
{M_1}=\sum\limits_{i = 1}^{{M_u}} {\log \left( {\frac{P({\lVert {{{{\bf{\tilde h}}}_1}} \rVert-{\varepsilon _1})^2}}{{(1 + \Delta i){\epsilon _b}}}} \right)}.
\end{equation}

Regarding the inner fractional SDP problem (\ref{op9.2}), for each bisection iteration, the computational complexity comes from solving a feasibility problem with LMI constraints. This feasibility problem involves $2K$ LMI constraints of size $N_t+1$, 3 LMI constraints of size $N_t$ and $2K$ LMI constraints of size 1. If the standard IPM is used, the arithmetic computation cost of solving such a problem should be on the order of $\ln{(1/\epsilon)}\sqrt \gamma  \zeta$, where $\gamma$ and $\zeta$ is given in (\ref{wc.order}). This fact completes the proof.

\bibliography{PHYSI_Journal}

\begin{thebibliography}{10}
\providecommand{\url}[1]{#1}
\csname url@samestyle\endcsname
\providecommand{\newblock}{\relax}
\providecommand{\bibinfo}[2]{#2}
\providecommand{\BIBentrySTDinterwordspacing}{\spaceskip=0pt\relax}
\providecommand{\BIBentryALTinterwordstretchfactor}{4}
\providecommand{\BIBentryALTinterwordspacing}{\spaceskip=\fontdimen2\font plus
\BIBentryALTinterwordstretchfactor\fontdimen3\font minus
  \fontdimen4\font\relax}
\providecommand{\BIBforeignlanguage}[2]{{%
\expandafter\ifx\csname l@#1\endcsname\relax
\typeout{** WARNING: IEEEtran.bst: No hyphenation pattern has been}%
\typeout{** loaded for the language `#1'. Using the pattern for}%
\typeout{** the default language instead.}%
\else
\language=\csname l@#1\endcsname
\fi
#2}}
\providecommand{\BIBdecl}{\relax}
\BIBdecl

\bibitem{jindal2006capacity}
N.~Jindal and Z.-Q. Luo, ``Capacity limits of multiple antenna multicast,'' in
  \emph{Proc. {IEEE} Int. Symp. Inf. Theory}, Seattle, WA, Jul. 2006, pp.
  1841--1845.

\bibitem{sidiropoulos2006transmit}
N.~D. Sidiropoulos, T.~N. Davidson, and Z.-Q.~T. Luo, ``Transmit beamforming
  for physical-layer multicasting,'' \emph{{IEEE} Trans. Signal Process.},
  vol.~54, no.~6, pp. 2239--2251, Jun. 2006.

\bibitem{kim2011optimal}
H.~Kim, D.~J. Love, and S.~Y. Park, ``Optimal and successive approaches to
  signal design for multiple antenna physical layer multicasting,''
  \emph{{IEEE} Trans. Commun.}, vol.~59, no.~8, pp. 2316--2327, Aug. 2011.

\bibitem{zhu2012precoder}
H.~Zhu, N.~Prasad, and S.~Rangarajan, ``Precoder design for physical layer
  multicasting,'' \emph{{IEEE} Trans. Signal Process.}, vol.~60, no.~11, pp.
  5932--5947, Nov. 2012.

\bibitem{lee2013a}
W.~Lee, H.~Park, H.-B. Kong, J.~S. Kwak, and I.~Lee, ``A new beamforming design
  for multicast systems,'' \emph{{IEEE} Trans. Veh. Technol.}, vol.~62, no.~8,
  pp. 4093--4097, Oct. 2013.

\bibitem{wu2013physical}
S.~X. Wu, W.-K. Ma, and A.~M.-C. So, ``Physical-layer multicasting by
  stochastic transmit beamforming and {A}lamouti space-time coding,''
  \emph{{IEEE} Trans. Signal Process.}, vol.~61, no.~17, pp. 4230--4245, Sep.
  2013.

\bibitem{du2013optimum}
B.~Du, Y.~Jiang, X.~Xu, and X.~Dai, ``Optimum beamforming for {MIMO}
  multicasting,'' \emph{EURASIP J. Adv. Signal Process.}, vol. 2013, no. 121,
  pp. 1--15, Dec. 2013.

\bibitem{shiu2011physical}
Y.-S. Shiu, S.~Y. Chang, H.-C. Wu, S.~C.-H. Huang, and H.-H. Chen, ``Physical
  layer security in wireless networks: a tutorial,'' \emph{{IEEE} Wireless
  Commun.}, vol.~18, no.~2, pp. 66--74, Apr. 2011.

\bibitem{he2013wireless}
\BIBentryALTinterwordspacing
B.~He, X.~Zhou, and T.~D. Abhayapala, ``Wireless physical layer security with
  imperfect channel state information: A survey,'' Jun. 2013. [Online].
  Available: \url{http://arxiv.org/abs/1307.4146}
\BIBentrySTDinterwordspacing

\bibitem{hong2013enhancing}
Y.-W.~P. Hong, P.-C. Lan, and C.-C.~J. Kuo, ``Enhancing physical-layer secrecy
  in multiantenna wireless systems: An overview of signal processing
  approaches,'' \emph{{IEEE} Signal Process. Mag.}, vol.~30, no.~5, pp. 29--40,
  Sep. 2013.

\bibitem{mukherjee2014principles}
A.~Mukherjee, S.~A. Fakoorian, J.~Huang, A.~L. Swindlehurst \emph{et~al.},
  ``Principles of physical layer security in multiuser wireless networks: A
  survey,'' \emph{{IEEE} Commun. Surveys Tuts.}, vol.~16, no.~3, pp.
  1550--1573, Aug. 2014.

\bibitem{schaefer2015secure}
R.~F. Schaefer, H.~Boche, and H.~V. Poor, ``Secure communication under channel
  uncertainty and adversarial attacks,'' \emph{Proc. IEEE}, vol. 103, no.~10,
  pp. 1796--1813, Oct. 2015.

\bibitem{yener2015wireless}
A.~Yener and S.~Ulukus, ``Wireless physical-layer security: Lessons learned
  from information theory,'' \emph{Proc. IEEE}, vol. 103, no.~10, pp.
  1814--1825, Oct. 2015.

\bibitem{wang2015enhancing}
H.-M. Wang and X.-G. Xia, ``Enhancing wireless secrecy via cooperation: signal
  design and optimization,'' \emph{IEEE Commun. Mag.}, vol.~53, no.~12, pp.
  47--53, Dec. 2015.

\bibitem{liu2016physical}
Y.~Liu, H.-H. Chen, and L.~Wang, ``Physical layer security for next generation
  wireless networks: Theories, technologies, and challenges,'' \emph{{IEEE}
  Commun. Surveys Tuts.}, vol.~19, no.~1, pp. 347--376, 2017.

\bibitem{negi2005secret}
R.~Negi and S.~Goel, ``Secret communication using artificial noise,'' in
  \emph{IEEE Veh. Technol. Conf.}, vol.~62, no.~3, Sep. 2005, p. 1906.

\bibitem{liao2011qos}
W.-C. Liao, T.-H. Chang, W.-K. Ma, and C.-Y. Chi, ``Qo{S}-based transmit
  beamforming in the presence of eavesdroppers: {A}n optimized
  artificial-noise-aided approach,'' \emph{{IEEE} Trans. Signal Process.},
  vol.~59, no.~3, pp. 1202--1216, Mar. 2011.

\bibitem{li2013transmit}
Q.~Li, M.~Hong, H.-T. Wai, Y.-F. Liu, W.-K. Ma, and Z.-Q. Luo, ``Transmit
  solutions for {MIMO} wiretap channels using alternating optimization,''
  \emph{{IEEE} J. Sel. Areas Commun.}, vol.~31, no.~9, pp. 1714--1727, Sep.
  2013.

\bibitem{zheng2015multi}
T.-X. Zheng, H.-M. Wang, J.~Yuan, D.~Towsley, and M.~H. Lee, ``Multi-antenna
  transmission with artificial noise against randomly distributed
  eavesdroppers,'' \emph{{IEEE} Trans. Commun.}, vol.~63, no.~11, pp.
  4347--4362, Nov. 2015.

\bibitem{zheng2017safe}
T.-X. Zheng, H.-M. Wang, Q.~Yang, and M.~H. Lee, ``Safeguarding decentralized
  wireless networks using full-duplex jamming receivers,'' \emph{{IEEE} Trans.
  Wireless Commun.}, vol.~16, no.~1, pp. 278--292, Jan. 2017.

\bibitem{liu2009secrecy}
R.~Liu and H.~V. Poor, ``Secrecy capacity region of a multi-antenna {G}aussian
  broadcast channel with confidential messages,'' \emph{{IEEE} Trans. Inf.
  Theory}, vol.~55, no.~3, pp. 1235--1249, Mar. 2009.

\bibitem{liu2010multiple}
R.~Liu, T.~Liu, H.~V. Poor, and S.~Shamai, ``Multiple-input multiple-output
  {G}aussian broadcast channels with confidential messages,'' \emph{{IEEE}
  Trans. Inf. Theory}, vol.~56, no.~9, pp. 4215--4227, Sep. 2010.

\bibitem{fakoorian2013on}
S.~A.~A. Fakoorian and A.~L. Swindlehurst, ``On the optimality of linear
  precoding for secrecy in the {MIMO} broadcast channel,'' \emph{{IEEE} J. Sel.
  Areas Commun.}, vol.~31, no.~9, pp. 1701--1713, Sep. 2013.

\bibitem{park2016weighted}
D.~Park, ``Weighted sum rate maximization of {MIMO} broadcast and interference
  channels with confidential messages,'' \emph{{IEEE} Trans. Wireless Commun.},
  vol.~15, no.~3, pp. 1742--1753, Mar. 2016.

\bibitem{park2016secrecy}
------, ``Secrecy sum rates of {MIMO} multi-receiver wiretap channels,''
  \emph{{IEEE} Commun. Lett.}, vol.~20, no.~9, pp. 1804--1807, Sep. 2016.

\bibitem{csiszar1978broadcast}
I.~Csisz{\'a}r and J.~K{\"o}rner, ``Broadcast channels with confidential
  messages,'' \emph{{IEEE} Trans. Inf. Theory}, vol.~24, no.~3, pp. 339--348,
  May 1978.

\bibitem{Hung2010Multiple}
H.~D. Ly, T.~Liu, and Y.~Liang, ``Multiple-input multiple-output {G}aussian
  broadcast channels with common and confidential messages,'' \emph{{IEEE}
  Trans. Inf. Theory}, vol.~56, no.~11, pp. 5477--5487, Oct. 2010.

\bibitem{liu2013new}
R.~Liu, T.~Liu, H.~V. Poor, and S.~Shamai~(Shitz), ``New results on
  multiple-input multiple-output broadcast channels with confidential
  messages,'' \emph{{IEEE} Trans. Inf. Theory}, vol.~59, no.~3, pp. 1346--1359,
  Mar. 2013.

\bibitem{ekrem2012capacity}
E.~Ekrem and S.~Ulukus, ``Capacity region of {G}aussian {MIMO} broadcast
  channels with common and confidential messages,'' \emph{{IEEE} Trans. Inf.
  Theory}, vol.~58, no.~9, pp. 5669--5680, Sep. 2012.

\bibitem{wyrembelski2014strong}
R.~F. Wyrembelski and H.~Boche, ``Robust broadcasting of common and
  confidential messages over compound channels: Strong secrecy and decoding
  performance,'' \emph{{IEEE} Trans. Inf. Forensics Security}, vol.~9, no.~10,
  pp. 1720--1732, Oct. 2014.

\bibitem{Wyrembelski2012Physical}
R.~Wyrembelski and H.~Boche, ``Physical layer integration of private, common,
  and confidential messages in bidirectional relay networks,'' \emph{{IEEE}
  Trans. Wireless Commun.}, vol.~11, no.~9, pp. 3170--3179, Sep. 2012.

\bibitem{Schaefer2014Physical}
R.~Schaefer and H.~Boche, ``Physical layer service integration in wireless
  networks: Signal processing challenges,'' \emph{{IEEE} Signal Process. Mag.},
  vol.~31, no.~3, pp. 147--156, Apr. 2014.

\bibitem{boyd2009convex}
S.~Boyd and L.~Vandenberghe, \emph{Convex optimization}.\hskip 1em plus 0.5em
  minus 0.4em\relax Cambridge, UK: Cambridge university press, 2009.

\bibitem{liang2009information}
Y.~Liang, H.~V. Poor \emph{et~al.}, ``Information theoretic security,''
  \emph{Foundations Trends Commun. Inf. Theory}, vol.~5, no. 4-5, pp. 355--580,
  Apr. 2009.

\bibitem{love2008overview}
D.~J. Love, R.~W. Heath~Jr, V.~K. Lau, D.~Gesbert, B.~D. Rao, and M.~Andrews,
  ``An overview of limited feedback in wireless communication systems,''
  \emph{{IEEE} J. Sel. Areas Commun.}, vol.~26, no.~8, pp. 1341--1365, Oct.
  2008.

\bibitem{Charnes1962Programming}
A.~Charnes and W.~W. Cooper, ``Programming with linear fractional
  functionals,'' \emph{Naval Res. Logist. Quart.}, vol.~9, no. 3-4, pp.
  181--186, 1962.

\bibitem{Boyd2011CVX}
\BIBentryALTinterwordspacing
M.~Grant and S.~Boyd, ``{CVX}: {M}atlab software for disciplined convex
  programming,'' Apr. 2011. [Online]. Available: \url{http://cvxr.com/cvx}
\BIBentrySTDinterwordspacing

\bibitem{Bertsekas1999}
D.~Bertsekas, \emph{Nonlinear Programming}, 2nd~ed.\hskip 1em plus 0.5em minus
  0.4em\relax Belmont, MA, USA: Athena Scientific, 1999.

\bibitem{ng2016multi}
D.~W.~K. Ng, E.~S. Lo, and R.~Schober, ``Multiobjective resource allocation for
  secure communication in cognitive radio networks with wireless information
  and power transfer,'' \emph{{IEEE} Trans. Veh. Technol.}, vol.~65, no.~5, pp.
  3166--3184, May 2016.

\bibitem{marler2004survey}
R.~T. Marler and J.~S. Arora, ``Survey of multi-objective optimization methods
  for engineering,'' \emph{Structural Multidisciplinary Optim.}, vol.~26,
  no.~6, pp. 369--395, Apr. 2004.

\bibitem{Huang2010Rank}
Y.~Huang and D.~Palomar, ``Rank-constrained separable semidefinite programming
  with applications to optimal beamforming,'' \emph{{IEEE} Trans. Signal
  Process.}, vol.~58, no.~2, pp. 664--678, Sep. 2010.

\bibitem{huang2012robust}
J.~Huang and A.~L. Swindlehurst, ``Robust secure transmission in {MISO}
  channels based on worst-case optimization,'' \emph{{IEEE} Trans. Signal
  Process.}, vol.~60, no.~4, pp. 1696--1707, Dec. 2012.

\bibitem{yoo2006capacity}
T.~Yoo and A.~Goldsmith, ``Capacity and power allocation for fading {MIMO}
  channels with channel estimation error,'' \emph{{IEEE} Trans. Inf. Theory},
  vol.~52, no.~5, pp. 2203--2214, May 2006.

\bibitem{shenouda2007convex}
M.~B. Shenouda and T.~N. Davidson, ``Convex conic formulations of robust
  downlink precoder designs with quality of service constraints,'' \emph{IEEE
  J. Sel. Topics Signal Process.}, vol.~1, no.~4, pp. 714--724, Dec. 2007.

\bibitem{li2013spatially}
Q.~Li and W.-K. Ma, ``Spatially selective artificial-noise aided transmit
  optimization for {MISO} multi-{E}ves secrecy rate maximization,''
  \emph{{IEEE} Trans. Signal Process.}, vol.~61, no.~10, pp. 2704--2717, May
  2013.

\bibitem{ben2001lectures}
A.~Ben-Tal and A.~Nemirovski, \emph{Lectures on modern convex optimization:
  analysis, algorithms, and engineering applications}.\hskip 1em plus 0.5em
  minus 0.4em\relax Philadelphia, PA, USA: SIAM, 2001, vol.~2.

\bibitem{horn2012matrix}
R.~A. Horn and C.~R. Johnson, \emph{Matrix analysis}.\hskip 1em plus 0.5em
  minus 0.4em\relax Cambridge, U.K.: Cambridge university press, 2012.

\end{thebibliography}
\bibliographystyle{IEEEtran}

\end{document}